\documentstyle[aps,epsfig]{revtex}

\def\S12{\mathrm{S}_{1,2}}
\newcommand{\ho}{\protect\mbox{$h^0$}}

\newcounter{allequation}

\begin{document}
\title{${}^{}$\\
[1.5cm] Is there a light fermiophobic Higgs ?\\
[1.5cm]}
\author{A. Barroso${}^{1}$}
\address{Dept. de F\'\i sica, Faculdade de Ci\^encias, Universidade de Lisboa\\
Campo Grande, C1, 1700 Lisboa, Portugal\\
and\\
CFNUL, Av. Prof. Gama Pinto 2, 1699 Lisboa, Portugal}
\author{L. Br\"ucher${}^{2}$}
\address{CFNUL, Av. Prof. Gama Pinto 2, 1699 Lisboa, Portugal}
\author{R. Santos${}^{3}$}
\address{Instituto Superior de Transportes, Campus Universit\'ario\\
R. D. Afonso Henriques, 2330 Entroncamento, Portugal\\
and\\
CFNUL, Av. Prof. Gama Pinto 2, 1699 Lisboa, Portugal\\
[2.5cm]}
\date{April 1999}
\maketitle

\begin{abstract}
The most general Two Higgs Doublet Model potential without explicit CP
violation depends on 10 real independent parameters. There are two different
ways of restricting this potential to 7 independent parameters. This gives
rise to two different potentials, $V_{(A)}$ and $V_{(B)}$. The phenomenology
of the two models is different, because some trilinear and quartic Higgs
couplings are different. As an illustration, we calculate the decay width of 
$h^0 \rightarrow \gamma \gamma$, where precisely due to the different
trilinear couplings the loop of the charged Higgs gives different
contributions. We also discuss the possibility for the existence of a light
fermiophobic Higgs.
\end{abstract}

\footnotetext[1]{
e-mail: barroso@alf1.cii.fc.ul.pt} 
\footnotetext[2]{
e-mail: bruecher@alf1.cii.fc.ul.pt} 
\footnotetext[3]{
e-mail: rsantos@alf1.cii.fc.ul.pt}



\newpage

\section{Introduction}

Despite the great success of the standard $SU(2)\times U(1)$ electroweak
model (SM), one of its fundamental principles, the spontaneous symmetry
breaking mechanism, still awaits experimental confirmation. This mechanism,
in its minimal version, requires the introduction of a single doublet of
scalar complex fields and gives rise to the existence of a neutral particle
with mass $m_H$. The combined analysis \cite{vanc} of all electroweak data
as a function of $m_H$ favors a value of $m_H$ close to $100\;GeV/c^2$ and
predicts with 95\% confidence level an upper bound of $m_H < 200 \;GeV/c^2$.
Hence, one can still envisage the possibility of a Higgs discovery in the
closing stages of the LEP operation.

Nevertheless, even if this turned out to be true, one still would like to
know if there is just one family of Higgs fields or, on the contrary, if
nature has decided to replicate itself. In our view this is the main
motivation to consider multi Higgs models. In this paper we continue the
study of the two-Higgs-doublet model (2HDM). Following our previous work 
\cite{Sant3}, we examine models without explicit CP violation and which are
also naturally protected from developing a spontaneous CP breaking minimum.
There are two different ways of achieving this. To illustrate the different
phenomenology we calculate, in both models, the decay width for the process $%
h^0 \rightarrow \gamma \gamma$, which can be particularly relevant if $h^0$
is a fermiophobic Higgs.

\section{The potentials}

The Higgs mechanism in its minimal version (one scalar doublet) introduces
in the theory an arbitrary parameter --- the Higgs boson mass $m_H$. In
fact, the potential depends on two parameters, which are the coefficients of
the quadratic and quartic terms. However the perturbative version of the
theory replaces them by the vacuum expectation value $v=247 \;GeV$ and by $%
m_H$. If we generalize the theory introducing a second doublet of complex
fields, the number of free parameters in the potential $V$ grows from two to
fourteen. At the same time, the number of scalar particles grows from one to
four. In this general form the potential contains genuine new interaction
vertices which are independent of the vacuum expectation values and of the
mass matrix of the Higgs bosons. However, these new interactions can be
avoided if one imposes the restriction that $V$ is invariant under charge
conjugation $C$. In fact, if $\Phi_i$ with $i=1,2$ denote two complex scalar
doublets with hyper-charge 1, under $C$ the fields transform themselves as $%
\Phi_i\rightarrow \exp(i\alpha_i) \Phi_i^*$ where the parameters $\alpha_i$
are arbitrary. Then, choosing $\alpha_1=\alpha_2=0$, and defining $x_1 =
\phi_1^\dagger \phi_1$, $x_2 = \phi_2^\dagger \phi_2$, $x_3 =
\Re\{\phi_1^\dagger \phi_2\}$ and $x_4 = \Im\{\phi_1^\dagger \phi_2\}$ it is
easy to see that the most general 2HDM potential without explicit $C$
violation\footnote{%
At this level $C$ conservation is equivalent to $CP$ conservation since all
fields are scalars.}, is: 
\begin{equation}  \label{pot}
V=-\mu_{1}^{2} x_{1}-\mu_{2}^{2} x_{2}-\mu_{12}^{2} x_{3}+\lambda_{1}
x_{1}^2+\lambda_{2} x_{2}^2+\lambda_{3} x_{3}^2+\lambda_{4}
x_{4}^2+\lambda_{5} x_{1} x_{2} +\lambda_6 x_1 x_3 +\lambda_7 x_2 x_3 %
\enskip .
\end{equation}

In general, the minimum of this potential is of the form

\setcounter{allequation}{\value{equation}} \addtocounter{allequation}{1} %
\setcounter{equation}{0} \renewcommand{\theequation}{\arabic{allequation}%
\alph{equation}} 
\begin{eqnarray}  \label{Ofields}
\left<\Phi_1\right> & = & \frac{1}{\sqrt{2}} \left(
\begin{array}{l}
0 \\ 
v_1
\end{array}
\right) \\
\left<\Phi_2\right> & = & \frac{1}{\sqrt{2}} \left(
\begin{array}{l}
0 \\ 
v_2 e^{i\theta}
\end{array}
\right) \quad ,
\end{eqnarray}
\setcounter{equation}{\value{allequation}} \renewcommand{\theequation}{%
\arabic{equation}} in other words it breaks $CP$ spontaneously. To use this
potential in perturbative electroweak calculations the physical parameters
that should replace the $\lambda$'s and $\mu$'s are the following:

\begin{enumerate}
\item[i)]  the position of the minimum, $v_{1}$, $v_{2}$ and $\theta $, or
alternatively, $v^{2}=v_{1}^{2}+v_{2}^{2}$, $\tan \beta =\frac{v_{2}}{v_{1}}$
and $\theta $;

\item[ii)]  the masses of the charged boson $m_{+}$ and of the three neutral
bosons $m_{1}$, $m_{2}$ and $m_{3}$;

\item[iii)]  and the three Cabibbo like angles $\alpha _{1}$, $\alpha _{2}$
and $\alpha _{3}$ that represent the orthogonal transformation that
diagonalizes the $3\times 3$ mass matrix\footnote{%
The mass matrix corresponding to the neutral components $\left( T_3 = - 
\frac{1}{2} \right)$ of the doublets is a $4\times 4$ matrix, but one
eigenvalue is zero because it corresponds to the $Z$ would be Goldstone
boson.} of the neutral sector.
\end{enumerate}

In a previous paper\cite{Sant3} we have examined the different types of
extrema for potential $V$. In particular it was shown in \cite{Sant3} that
there are two ways of naturally imposing that a minimum with $CP$ violation
never occurs. This, in turn, leads to two different 7-parameter potentials.
The first one, denoted $V_{(A)}$, is the potential discussed in the review
article of M. Sher\cite{sher1} and corresponds to setting $%
\mu_{12}^2=\lambda_6=\lambda_7=0$ in equation (\ref{pot}). The second
7-parameter potential, that we shall call $V_{(B)}$, is essentially the
version analyzed in the Higgs Hunters Guide\cite{Gun} and it corresponds to
the conditions $\lambda_6=\lambda_7=0$ and $\lambda_3=\lambda_4$. As we have
already pointed out \cite{Sant3} but would like to stress again, these
potentials have different phenomenology. This is illustrated in section \ref
{fermiophob} when we consider the fermiophobic limit of both models.

Since $V_{(A)}$ and $V_{(B)}$ do not have spontaneous $CP$-violation, the
number of so-called ``physical parameters'' is immediately reduced to seven.
In fact, $\theta=0$ and only one rotation angle, $\alpha$, is needed to
diagonalize the $2\times 2$ mass matrix of the $CP$-even neutral scalars.
This is clearly seen if we transform the initial doublets $\Phi_i$ into two
new ones $H_i$ given by 
\begin{equation}  \label{e:fieldrot}
\left(
\begin{array}{c}
H_1 \\ 
H_2
\end{array}
\right) = \frac{1}{\sqrt{v_1^2 + v_2^2}} \left(
\begin{array}{rr}
v_1 & v_2 \\ 
-v_2 & v_1
\end{array}
\right) \left(
\begin{array}{c}
\Phi_1 \\ 
\Phi_2
\end{array}
\right) \quad .
\end{equation}
In this Higgs bases, only $H_1$ acquires a vacuum expectation value. Then,
the $T_3=+\frac{1}{2}$  component and the imaginary part of the $T_3=-\frac{1%
}{2}$ component of $H_1$ are the $W^\pm$ and $Z$ would be Goldstone bosons,
respectively. The $C$-odd neutral boson, $A^0$, is the imaginary part of the 
$T_3=-\frac{1}{2}$ component of $H_2$. On the other hand, the light and
heavy $CP$-even neutral Higgs, $h^0$ and $H^0$, are linear combinations of
the real parts of the $T_3=-\frac{1}{2}$ component of $H_1$ and $H_2$.

Notice that $V_{(A)}$ is invariant under the $Z_2$ transformation $\Phi_1
\rightarrow \Phi_1$ and $\Phi_2 \rightarrow -\Phi_2$, whereas in $V_{(B)}$
only the $\mu_{12}^2$ term breaks the $U(1)$ symmetry, $\Phi_2 \rightarrow
e^{i \alpha}\Phi_2$. Because this breaking occurs in a quadratic term it
does not spoil the renormalizability of the model. Hence, in both cases the
terms that were set explicitly to zero, will not be needed to absorb
infinities that occur at higher orders. The complete renormalization program
of the model based on $V_{(A)}$ was carried out in \cite{Sant1}. The results
for $V_{(B)}$ are similar but the cubic and quartic scalar vertices have to
be changed appropriately.

For the sake of completeness we will close this section with a summary of
the results that will be used later. As we have already said they are not
new and can be obtained either from \cite{sher1} or \cite{Gun}. We agree
with both.

For $V_{(A)}$ the minimum conditions are \setcounter{allequation}{%
\value{equation}} \addtocounter{allequation}{1} \setcounter{equation}{0} %
\renewcommand{\theequation}{\arabic{allequation}\alph{equation}} 
\begin{eqnarray}  \label{vac1}
0 \quad = \quad T_1 & = & v_{{1}}\left (\,-\mu_{{1}}^2 + \lambda_{{1}}{v_{{1}%
}}^{2}+ \lambda_{+}\,{v_{{2}}}^{2}\right ) \\
0 \quad = \quad T_2 & = & v_{{2}}\left (\,-\mu_{{2}}^2 + \lambda_{{2}}{v_{{2}%
}}^{2}+ \lambda_{{+}}\,{v_{{1}}}^{2}\right )
\end{eqnarray}
\setcounter{equation}{\value{allequation}} \renewcommand{\theequation}{%
\arabic{equation}} with $\lambda_+=\frac{1}{2}(\lambda_3+\lambda_5)$. They
lead to the following solutions:\newline
either i) \setcounter{allequation}{\value{equation}} %
\addtocounter{allequation}{1} \setcounter{equation}{0} \renewcommand{%
\theequation}{\arabic{allequation}\alph{equation}} 
\begin{eqnarray}
v_1^2 & = & \frac{\lambda_2 \mu_1^2 - \lambda_+ \mu_2^2} {
\lambda_1\lambda_2 - \lambda_+^2 } \\
v_2^2 & = & \frac{\lambda_1 \mu_2^2 - \lambda_+ \mu_1^2} {
\lambda_1\lambda_2 - \lambda_+^2 } \quad ;
\end{eqnarray}
\setcounter{equation}{\value{allequation}} \renewcommand{\theequation}{%
\arabic{equation}} or\phantom{eth} ii) \setcounter{allequation}{%
\value{equation}} \addtocounter{allequation}{1} \setcounter{equation}{0} %
\renewcommand{\theequation}{\arabic{allequation}\alph{equation}} 
\begin{eqnarray}
v_1^2 & = & 0 \\
v_2^2 & = & \frac{\mu_2^2}{\lambda_2} \quad .
\end{eqnarray}
\setcounter{equation}{\value{allequation}} \renewcommand{\theequation}{%
\arabic{equation}} \newline
The masses of the Higgs bosons and the angle $\alpha$ are given by the
following relations: \setcounter{allequation}{\value{equation}} %
\addtocounter{allequation}{1} \setcounter{equation}{0} \renewcommand{%
\theequation}{\arabic{allequation}\alph{equation}} 
\begin{eqnarray}  \label{massrel1}
m_{H^+}^2 & = & -\lambda_3 \,\left( v_1^2 + v_2^2 \right) \\
m_{A^0}^2 & = & \frac{1}{2} \left(\lambda_4 - \lambda_3 \right) \,\left(
v_1^2 + v_2^2 \right) \\
m_{H^0,h^0}^2 & = & \lambda_1 v_1^2 + \lambda_2 v_2^2 \, \pm \, \sqrt{\left(
\lambda_1 v_1^2 - \lambda_2 v_2^2 \right)^2 + v_1^2 v_2^2 (\lambda_3 +
\lambda_5)^2 }
\end{eqnarray}
\setcounter{equation}{\value{allequation}} \renewcommand{\theequation}{%
\arabic{equation}} 
\begin{equation}  \label{alph1}
\tan 2\alpha = \frac{v_2 v_1 \,(\lambda_3 + \lambda_5)}{\lambda_1 v_1^2 -
\lambda_2 v_2^2} \qquad . \qquad\qquad\qquad\qquad\qquad\qquad\quad
\end{equation}

On the other hand, for $V_{(B)}$ the minimum conditions are %
\setcounter{allequation}{\value{equation}} \addtocounter{allequation}{1} %
\setcounter{equation}{0} \renewcommand{\theequation}{\arabic{allequation}%
\alph{equation}} 
\begin{eqnarray}  \label{vac2}
0 & = & T_1 - \frac{\mu_{12}^2}{2}\, v_2 \\
0 & = & T_2 - \frac{\mu_{12}^2}{2}\, v_1
\end{eqnarray}
\setcounter{equation}{\value{allequation}} \renewcommand{\theequation}{%
\arabic{equation}} with the $T_i$ given by the previous equations (4). The
solution of this set of equations is \setcounter{allequation}{%
\value{equation}} \addtocounter{allequation}{1} \setcounter{equation}{0} %
\renewcommand{\theequation}{\arabic{allequation}\alph{equation}} 
\begin{eqnarray}
v_1^2 & = & \frac{\lambda_1 -\lambda_2 \pm \sqrt{\left(\lambda_1
-\lambda_2\right)^2 - 4\left(\lambda_1 -\lambda_+ \right) \left(\lambda_2
-\lambda_+ \right)\left[\left(\lambda_+ v^2 - \mu_1^2 \right)
\left(\lambda_2 v^2 - \mu_2^2 \right)-\mbox{$\frac{1}{4}$} \mu_{12}^4
\right] }} {2\,\left(\lambda_1 -\lambda_+ \right)\left(\lambda_2
-\lambda_+\right)} \\
v_2^2 & = & \frac{\lambda_2 -\lambda_1 \pm \sqrt{\left(\lambda_1
-\lambda_2\right)^2 - 4\left(\lambda_2 -\lambda_+ \right) \left(\lambda_1
-\lambda_+ \right)\left[\left(\lambda_+ v^2 - \mu_2^2 \right)
\left(\lambda_1 v^2 - \mu_1^2 \right)-\mbox{$\frac{1}{4}$} \mu_{12}^4
\right] }} {2\,\left(\lambda_1 -\lambda_+ \right)\left(\lambda_2
-\lambda_+\right)} \quad .
\end{eqnarray}
\setcounter{equation}{\value{allequation}} \renewcommand{\theequation}{%
\arabic{equation}} Notice that, in this case, the solution with vanishing
vacuum expectation value in one of the doublets is not possible. Now the
masses and the value of $\alpha$ are given by \setcounter{allequation}{%
\value{equation}} \addtocounter{allequation}{1} \setcounter{equation}{0} %
\renewcommand{\theequation}{\arabic{allequation}\alph{equation}} 
\begin{eqnarray}  \label{massrel2c}
m_{H^+}^2 & = & -\lambda_3 \,\left( v_1^2 + v_2^2 \right)+\mu_{12}^2 \,\frac{%
v_1^2 + v_2^2}{v_1 v_2} \\
m_{A^0}^2 & = & \frac{1}{2} \mu_{12}^2 \,\frac{v_1^2 + v_2^2}{v_1 v_2} \\
m_{H^0,h^0}^2 & = & \lambda_1 v_1^2 + \lambda_2 v_2^2 + \mbox{$\frac{1}{4}$}
\mu_{12}^2\left(\mbox{$\frac{v_2}{v_1}$}+\mbox{$\frac{v_1}{v_2}$}\right)\, \\
& & \pm \, \sqrt{\left( \lambda_1 v_1^2 - \lambda_2 v_2^2 + %
\mbox{$\frac{1}{4}$} \mu_{12}^2\left(\mbox{$\frac{v_2}{v_1}$}-%
\mbox{$\frac{v_1}{v_2}$}\right) \right)^2 + \left( v_1 v_2 (\lambda_3 +
\lambda_5) - \mbox{$\frac{1}{2}$} \mu_{12}^2\right)^2 }  \nonumber
\end{eqnarray}
\setcounter{equation}{\value{allequation}} \renewcommand{\theequation}{%
\arabic{equation}} 
\begin{equation}  \label{alph2}
\tan 2\alpha = \frac{2\,v_1 v_2 \,\lambda_+ -\mbox{$\frac{1}{2}$} \mu_{12}^2%
}{\lambda_1 v_1^2 - \lambda_2 v_2^2 + \mbox{$\frac{1}{4}$} \mu_{12}^2 \left(%
\mbox{$\frac{v_2}{v_1}$}-\mbox{$\frac{v_1}{v_2}$}\right)} \qquad .
\qquad\qquad\qquad\qquad\qquad
\end{equation}

\section{The fermiophobic limit}

\label{fermiophob}

Despite the fact that $V_{(A)}$ and $V_{(B)}$ are different, it is obvious
that the gauge bosons and the fermions couplings to the scalars are the same
for both models. In particular, the introduction of the Yukawa couplings
without tree-level flavor changing neutral currents is easily done extending
the $Z_2$ symmetry to the fermions. This leads to two different ways of
coupling the quarks and two different ways of introducing the leptons,
giving a total of four different models, usually denoted as model I, II, III
and IV (cf. e.g. \cite{Sant1}).

In here, we use model I, where only $\Phi_2$ couples to the fermions. Then,
the coupling of the lightest scalar Higgs, $h^0$, to a fermion pair (quark
or lepton) is proportional to $\cos\alpha$. As $\alpha$ approaches $\frac{\pi%
}{2}$ this coupling tends to zero and in the limit it vanishes, giving rise
to a fermiophobic Higgs.

Examining equations (\ref{alph1}) and (\ref{alph2}) we see that the
fermiophobic limit ($\alpha=\frac{\pi}{2}$) can be obtained in potential A
in two ways: either $\lambda_+=0$ or $v_1=0$. In potential B there is only
one possibility $2 v_1 v_2 \lambda_+=\frac{1}{2} \mu_{12}^2$. In this latter
case, equations (11) and (\ref{alph2}) give immediately: %
\setcounter{allequation}{\value{equation}} \addtocounter{allequation}{1} %
\setcounter{equation}{0} \renewcommand{\theequation}{\arabic{allequation}%
\alph{equation}} 
\begin{eqnarray}  \label{B_mh0}
m_{A^0}^2 & = & 2\,\lambda_+ \,\left(v_1^2+v_2^2\right) \\
m_{H^0}^2 & = & 2\,\lambda_2 v_2^2 \,+\,2\,\lambda_+ v_1^2 \quad = \quad
m_{A^0}^2 \,+\,2(\lambda_2-\lambda_+)\,v^2\sin^2\beta \\
m_{h^0}^2 & = & 2\,\lambda_1 v_1^2\,+\,2\,\lambda_+ v_2^2 \quad = \quad
m_{A^0}^2\,-\,2(\lambda_+-\lambda_1)\,v^2\cos^2\beta \quad .
\end{eqnarray}
\setcounter{equation}{\value{allequation}} \renewcommand{\theequation}{%
\arabic{equation}} In the former case ($V_{(A)}$), $\lambda_+=0$ gives %
\setcounter{allequation}{\value{equation}} \addtocounter{allequation}{1} %
\setcounter{equation}{0} \renewcommand{\theequation}{\arabic{allequation}%
\alph{equation}} 
\begin{eqnarray}
m_{H^0}^2 & = & 2\,\lambda_2 v_2^2 \\
m_{h^0}^2 & = & 2\,\lambda_1 v_1^2
\end{eqnarray}
\setcounter{equation}{\value{allequation}} \renewcommand{\theequation}{%
\arabic{equation}} while $v_1=0$ gives a massless $h^0$. In this analysis we
have assumed that $v_1<v_2$. The reversed situation leads to similar
conclusions since one is then interchanging the role of the two doublets.

The triple couplings involving two gauge bosons and a scalar particle like,
for instance $Z_\mu Z^\mu h^0$, are always proportional to the angle $%
\delta=\alpha-\beta$. In particular, the couplings for $h^0$ are
proportional to $\sin\delta$ whereas the corresponding $H^0$ couplings are
proportional to $\cos\delta$. This general results can be understood if one
recalls the argument about the role played by the neutral scalars in
restoring the unitarity in the scattering of longitudinal $W$'s, i.e. in $%
W^+_L W^-_L \rightarrow W^+_L W^-_L$. The restoration of unitarity requires
that the sum of the squares of the $W^+W^-h^0$ and $W^+W^-H^0$ couplings
adds up to a constant proportional to the $SU(2)$ gauge coupling, $g$.

Current searches of the SM Higgs boson at LEP put the mass limit at $89
\;GeV/c^2$\cite{lepweg}. Since the production mechanism is the reaction $e^+
e^- \rightarrow Z^* \rightarrow Z h^0$, this limit can be substantially
lower in the 2HDM if $\sin\delta$ is small. In our numerical application to
the two $\gamma$ decay of a light fermiophobic $h^0$ we will explore the
region $\sin^2\delta \le 0.1$ \cite{kraw1}.

Bounds on the Higgs masses have been derived by several authors \cite{Aker1}%
. Recently next-to-leading order calculations \cite{ciu1} in the SM give a
prediction for the branching ratio $Br(B\rightarrow X_s\gamma)$ which is
slightly larger than the experimental CLEO measurement\cite{cleo1}. In model
II the charged Higgs loops always increase the SM value. Hence, this process
provides good lower bounds on $m_{H^\pm}$ as a function of $\tan\beta$\cite
{ciu1}. On the contrary, in model I the contribution from the charged Higgs
reduces the theoretical prediction and so brings it to a value closer to the
experimental result. This reduction is larger for small $\tan\beta$, since
in model I the $H^+$ coupling to quarks is proportional to $\tan^{-1}\beta$.
However, a small $\tan\beta$ gives a large top Yukawa coupling which leads
to large new contributions to $R_b$, the $B_0-\bar{B}_0$ mixing. A recent
analysis by Ciuchini et al. \cite{ciu1} derives the bounds $\tan\beta >
1.8,\ 1.4 \mbox{ and } 1.0$ for $m_{H^\pm}=85,\ 200 \mbox{ and } 425
\;GeV/c^2$, respectively.

The Higgs contribution to the $\rho $-parameter is \cite{den1}: 
\begin{equation}
\Delta \rho =\frac{1}{16\pi ^{2}v^{2}}\left[ \sin ^{2}\delta \;F(m_{H^{\pm
}}^{2},m_{A^{0}}^{2},m_{H^{0}}^{2})\,+\,\cos ^{2}\delta \;F(m_{H^{\pm
}}^{2},m_{A^{0}}^{2},m_{h^{0}}^{2})\right]   \label{rho}
\end{equation}
where 
\[
F(a,b,c)=a\,+\,\frac{bc}{b-c}\ln \frac{b}{c}\,-\,\frac{ab}{a-b}\ln \frac{a}{b%
}\,-\,\frac{ac}{a-c}\ln \frac{a}{c}\quad .
\]
Since the current experimental value of $\rho =1.0012\pm 0.0013\pm 0.0018$ 
\cite{Rev} exceeds the SM prediction by 3$\sigma $, one should at least try
to avoid a positive $\Delta \rho $.\footnote{%
A more recent SM fit gives $\rho =0.9996+0.0031(-0.0013)\cite{Lang98}.$}. A
simpler examination of the function $F(a,b,c)$ shows that this is impossible
if $m_{H^{\pm }}$ is the largest mass. On the other hand, if $%
m_{A^{0}}>m_{H^{\pm }}$ one obtains a negative value for $\Delta \rho $
which grows with the splitting $m_{A^{0}}-m_{H^{\pm }}$. In line with our
limit ($\sin ^{2}\delta \le 0.1$), negative values of $\Delta \rho $ of the
order of the experimental statistical error, i.e. $\Delta \rho \approx
-10^{-3}$, can be obtained essentially in two ways. Either with a large $%
m_{H^{\pm }}\approx 300\;GeV/c^{2}$ but with a modest $m_{A^{0}}-m_{H^{\pm }}
$ splitting ($m_{A^{0}}\approx 340\;GeV/c^{2}$) or with a smaller $m_{H^{\pm
}}\approx 100\;GeV/c^{2}$ but with $m_{A^{0}}\approx 200\;GeV/c^{2}$. The
variation of $\Delta \rho $ with $m_{h^{0}}$ is rather modest, less than $10$%
\% for the range $20\;GeV/c^{2}\le m_{h^{0}}\le 100\;GeV/c^{2}$. With seven
parameters in the Higgs sector it is difficult and not very illuminating to
discuss in detail all possibilities. So, this discussion should be regarded
as a simple justification for the fact that a fermiophobic Higgs scenario is
not ruled out by the existing experiments. We would like to stress, that
there could exist a light $h^{0}$ almost decoupled from the fermions ($%
\alpha \approx \frac{\pi }{2}$) and at the same time with a small LEP
production rate via the $Z$-bremsstrahlungs reaction $Z^{*}\rightarrow
Zh^{0}\;(\sin ^{2}\delta \approx 10^{-1})$. If such a boson exists it will
decay mainly via the process $h^{0}\rightarrow \gamma \gamma $.

\section{The decay $\ho \longrightarrow \gamma \gamma$}

The decay $\mbox{$h^0$} \rightarrow \gamma \gamma$ is particularly suitable
to illustrate the fact that $V_{(A)}$ and $V_{(B)}$ give rise to different
phenomenologies. In fact, the decay occurs at one-loop level and for a
fermiophobic Higgs one has vector bosons and charged Higgs contributions.
The latter are different for models $A$ and $B$, because the $h^0 H^+ H^-$
vertex is different. It is interesting to point out how this difference
arises. Since the term in $\lambda_4$ does not contribute to this vertex,
both potentials give rise to the same effective $h^0 H^+ H^-$ coupling, $%
g_{h^0 H^+ H^-}$, namely:

\begin{eqnarray}  \label{e:c_org_A}
[h^0 H^+ H^-] & = & 2\,v_2\lambda_2\cos^2\beta \cos\alpha + v_2\lambda_3
\sin\alpha \cos\beta \sin\beta - v_1\lambda_5\cos^2\beta \sin\alpha \\
& & - 2\,v_1\lambda_1 \sin^2\beta \sin\alpha + v_2\lambda_5\sin^2\beta
\cos\alpha - v_1\lambda_3\cos\alpha\cos\beta\sin\beta  \nonumber
\end{eqnarray}

However, as we have already said, what is relevant for perturbative
calculations is the position of the minimum of $V$ and the values of its
derivatives at that point. This means that one has to express all coupling
constants in terms of the particle masses. This is simply done by inverting
equations (7) and (11). The result is 
\begin{equation}  \label{coup1}
[h^0 H^+ H^-]_{(A)} = \frac{g}{m_W} \left( m_{h^0}^2 \frac{\cos\left(
\alpha\!+\!\beta\right)}{\sin 2\beta} - \left(m_{H^+}^2-\mbox{$\frac{1}{2}$}
m_{h^0}^2\right)\sin \left( \alpha\!-\!\beta\right) \,\right)
\end{equation}
and 
\begin{equation}  \label{coup2}
[h^0 H^+ H^-]_{(B)} = \frac{g}{m_W} \left( \left(m_{h^0}^2 -m_{A^0}^2\right)%
\frac{\cos\left( \alpha\!+\!\beta\right)}{\sin 2\beta} - \left(m_{H^+}^2-%
\mbox{$\frac{1}{2}$} m_{h^0}^2\right)\sin \left( \alpha\!-\!\beta\right)
\,\right)
\end{equation}
which clearly shows the difference that we have pointed out.

\begin{figure}[htbp]
  \begin{center}
\epsfig{file=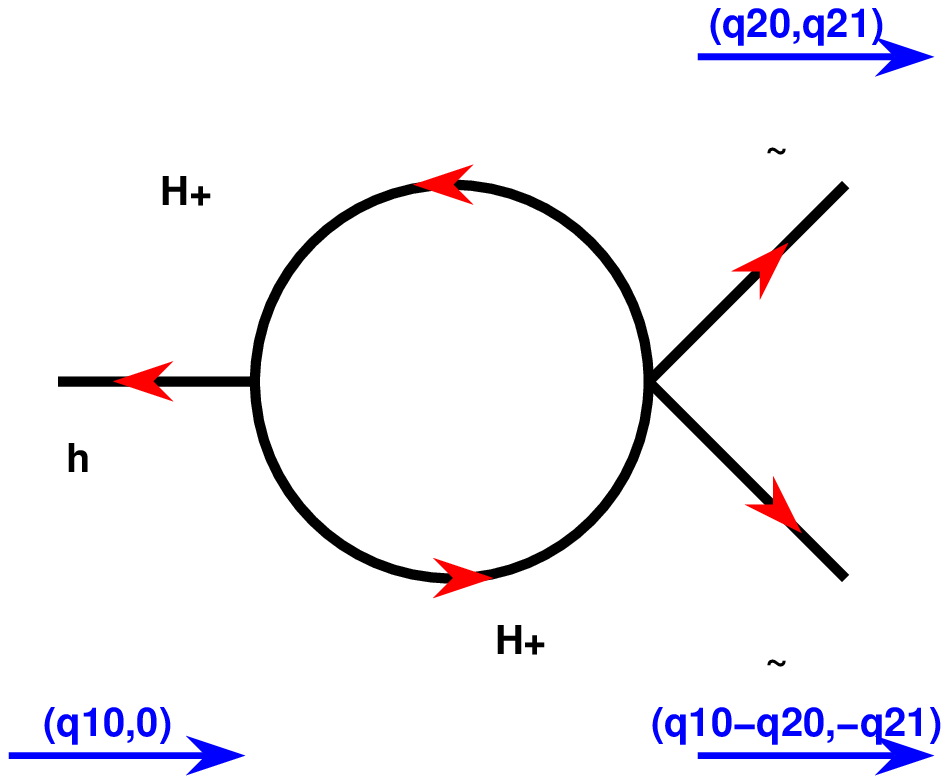,width=3.7cm}
\epsfig{file=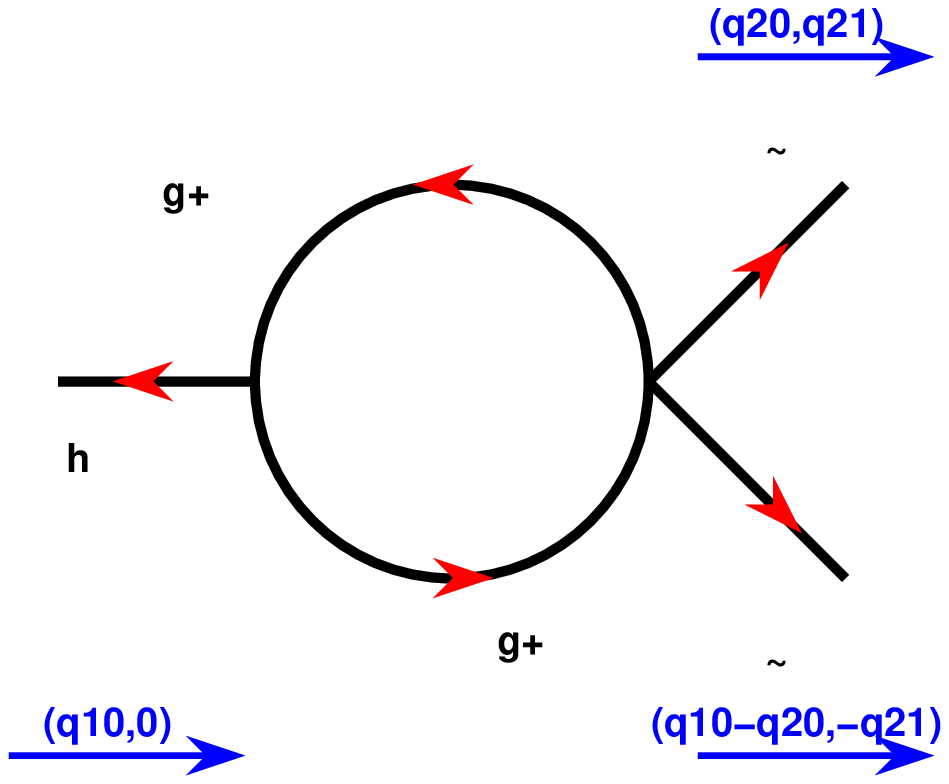,width=3.7cm}
\epsfig{file=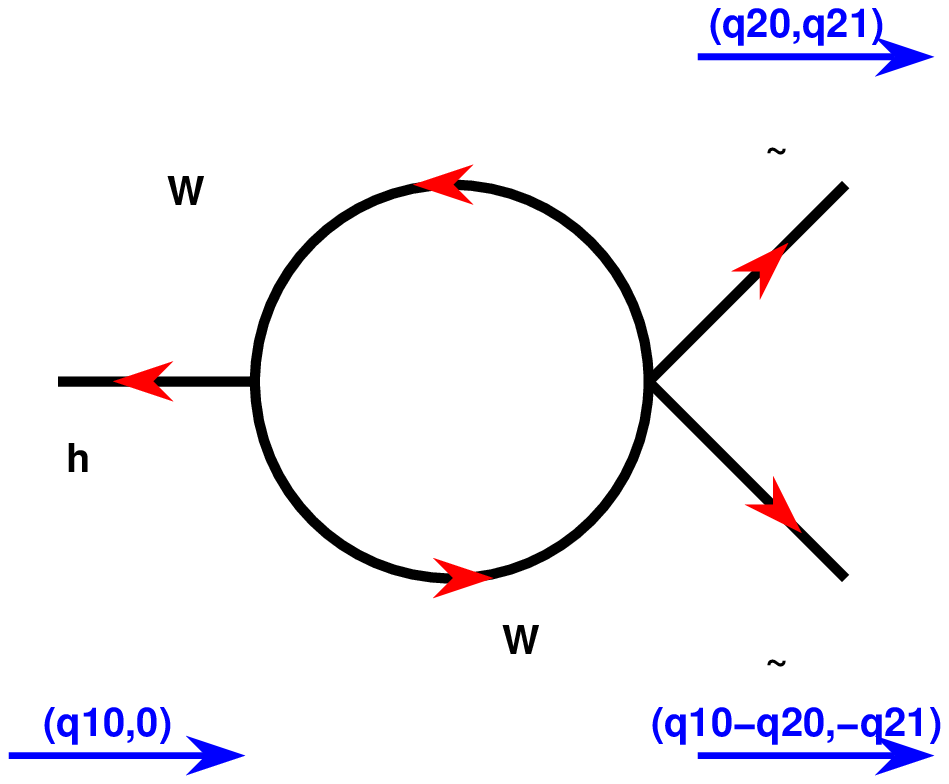,width=3.7cm}
\epsfig{file=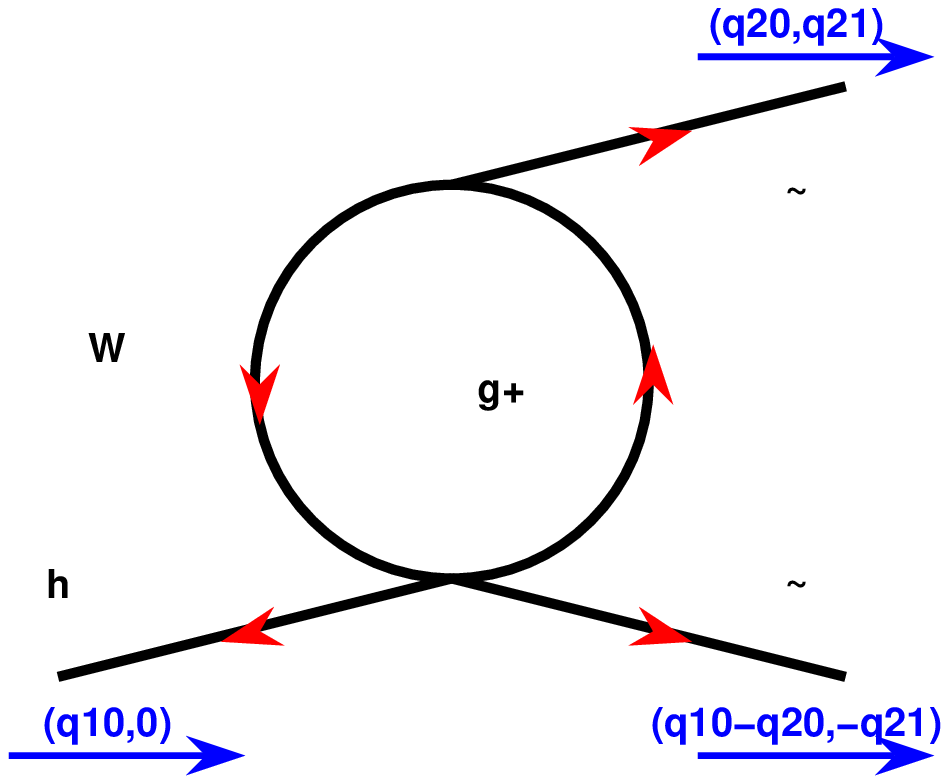,width=3.7cm}\\[0.1cm]
\epsfig{file=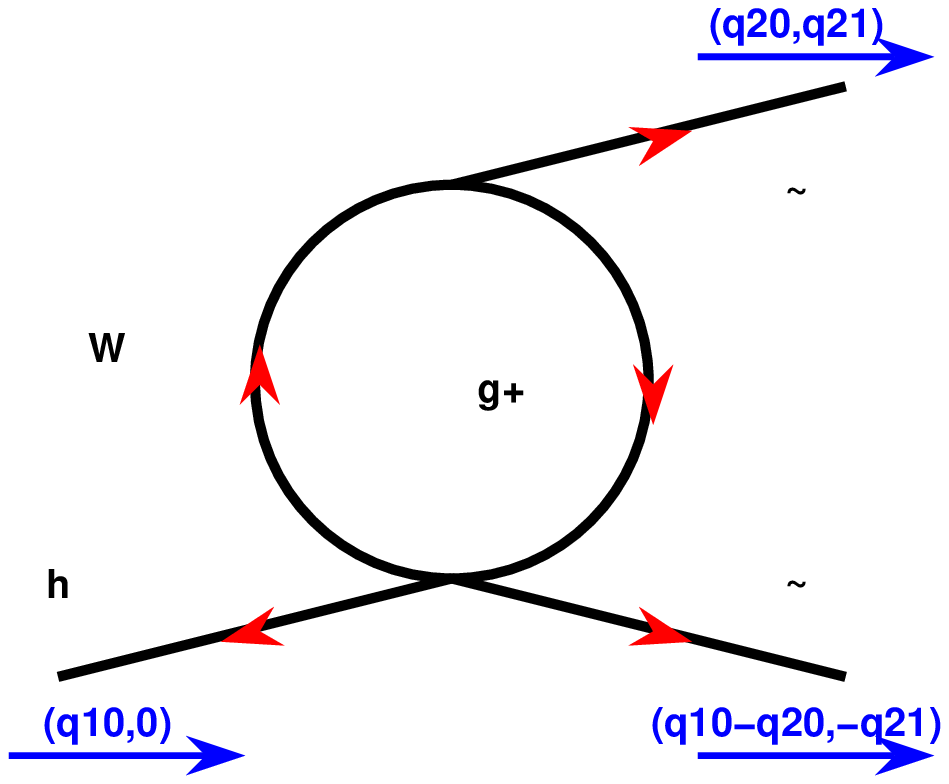,width=3.7cm}
\epsfig{file=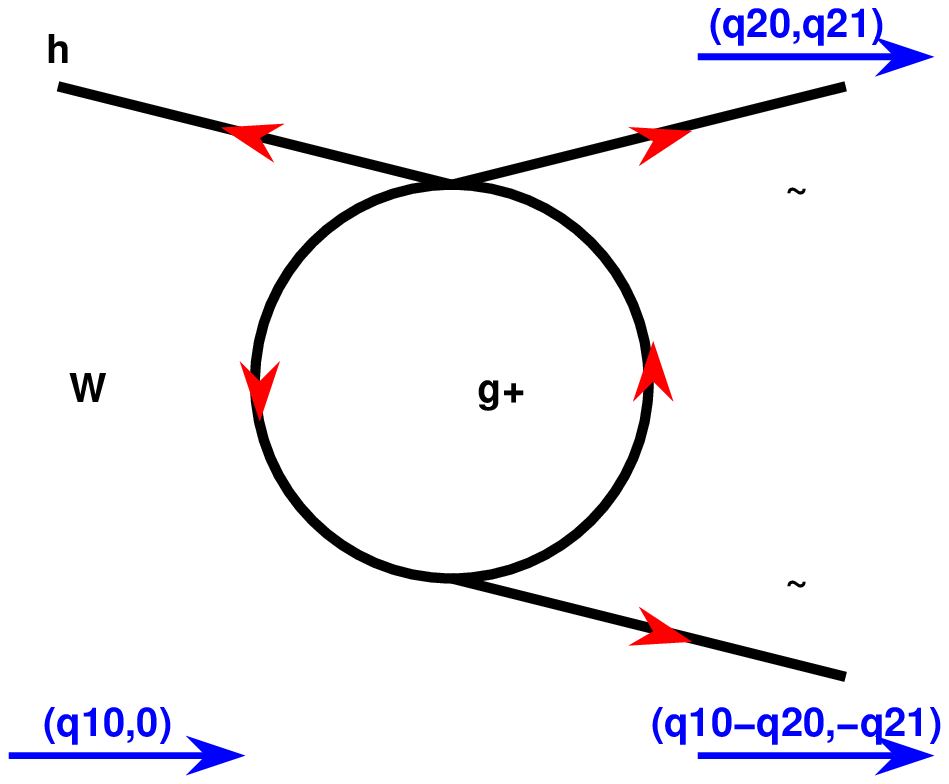,width=3.7cm}
\epsfig{file=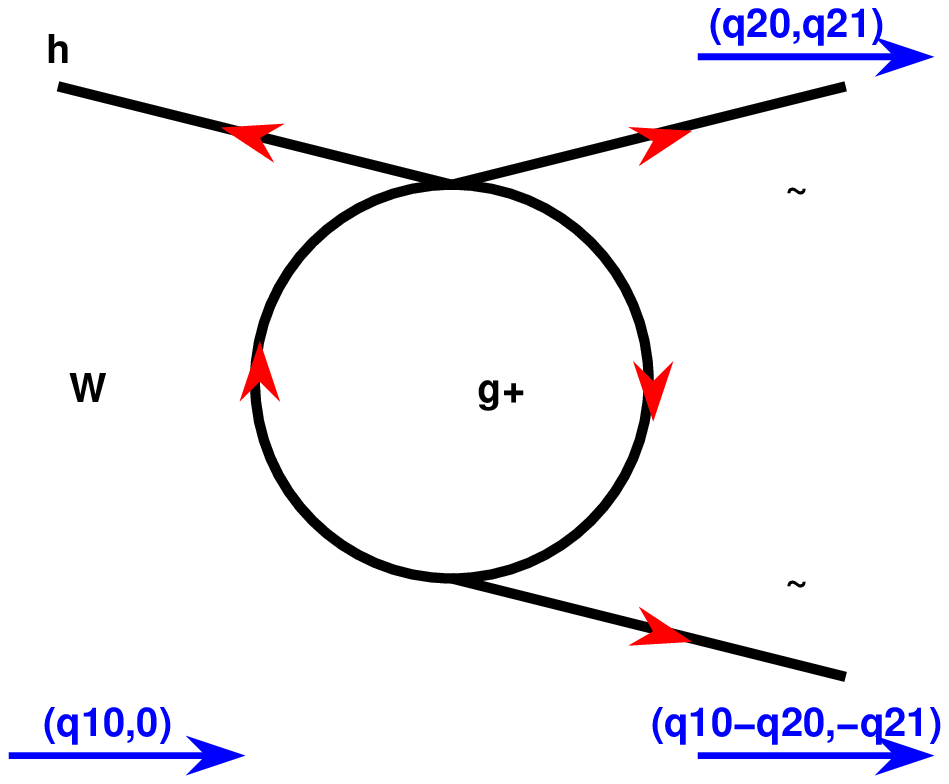,width=3.7cm}
\epsfig{file=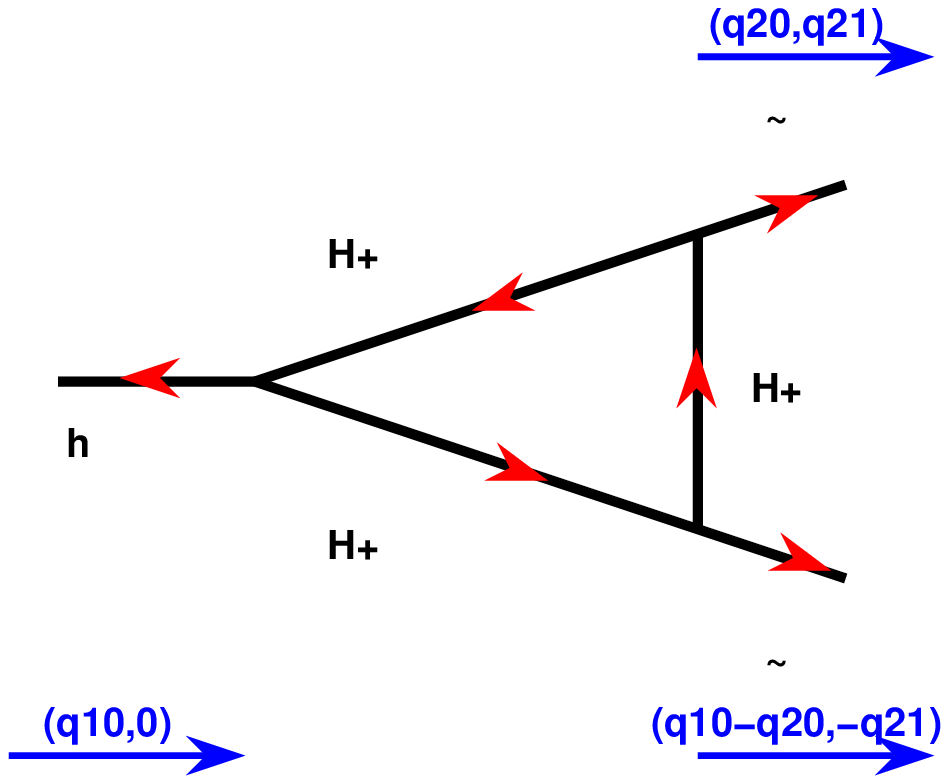,width=3.7cm}\\[0.1cm]
\epsfig{file=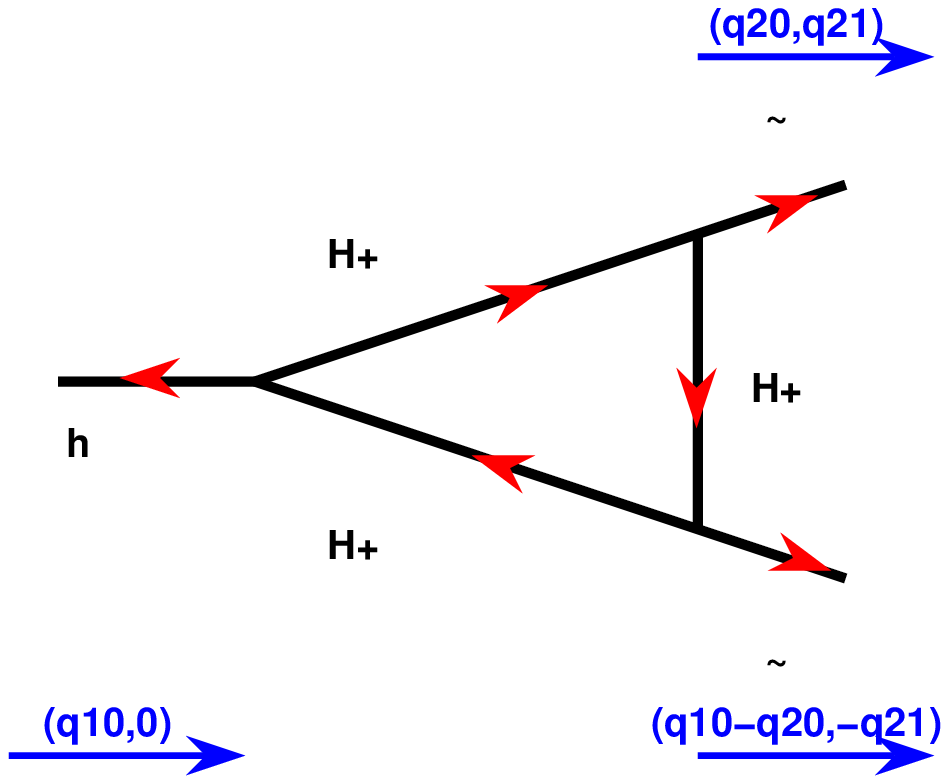,width=3.7cm}
\epsfig{file=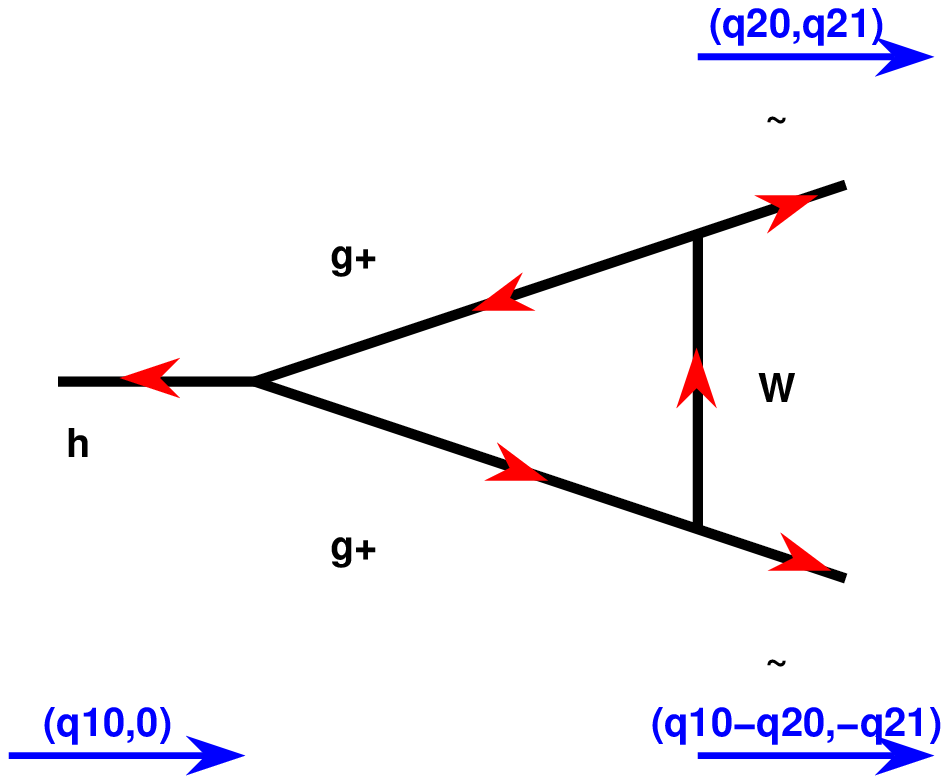,width=3.7cm}
\epsfig{file=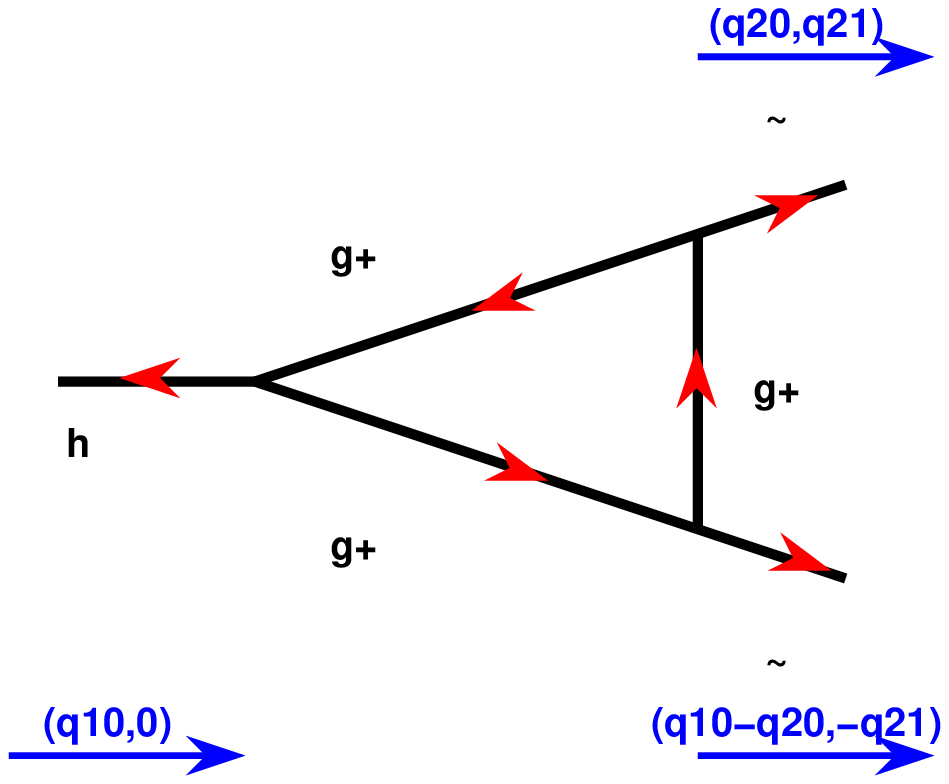,width=3.7cm}
\epsfig{file=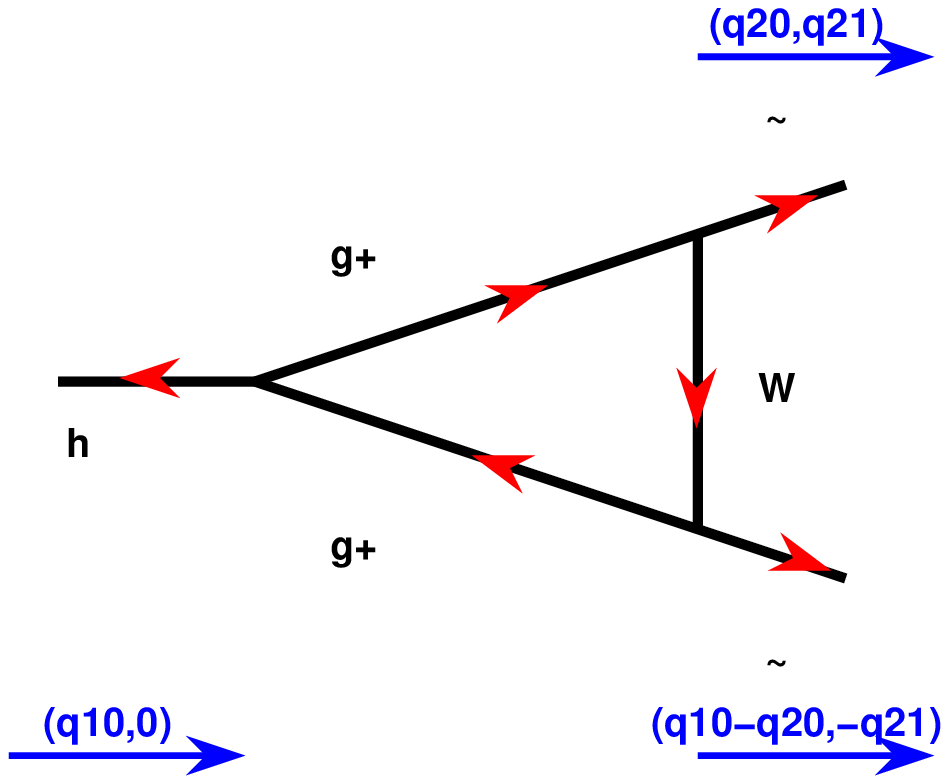,width=3.7cm}\\[0.1cm]
\epsfig{file=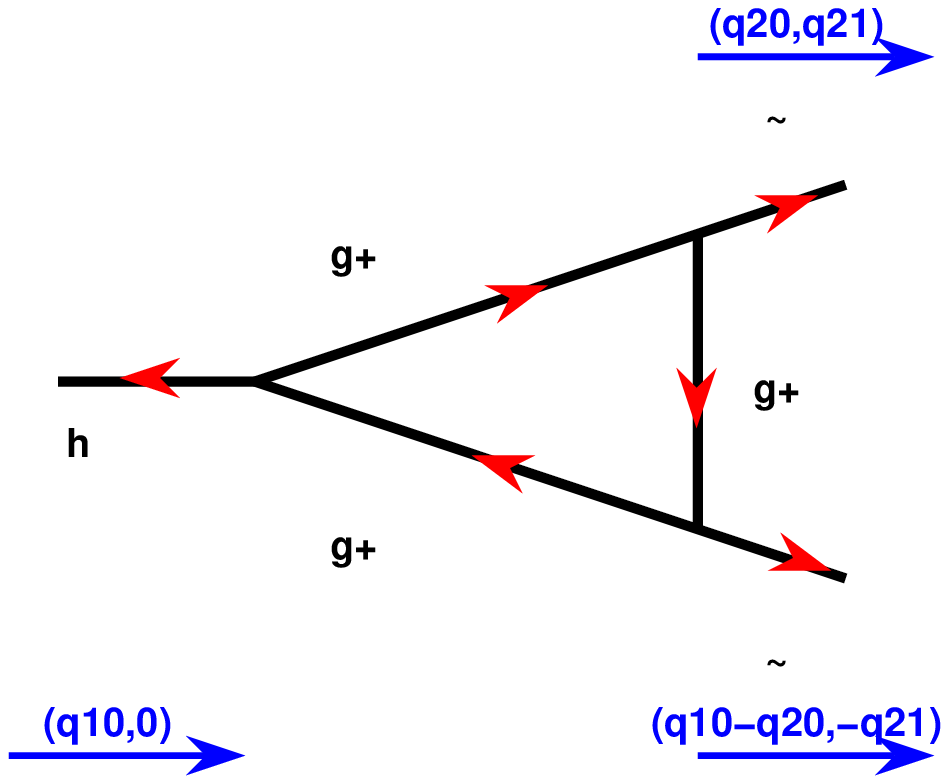,width=3.7cm}
\epsfig{file=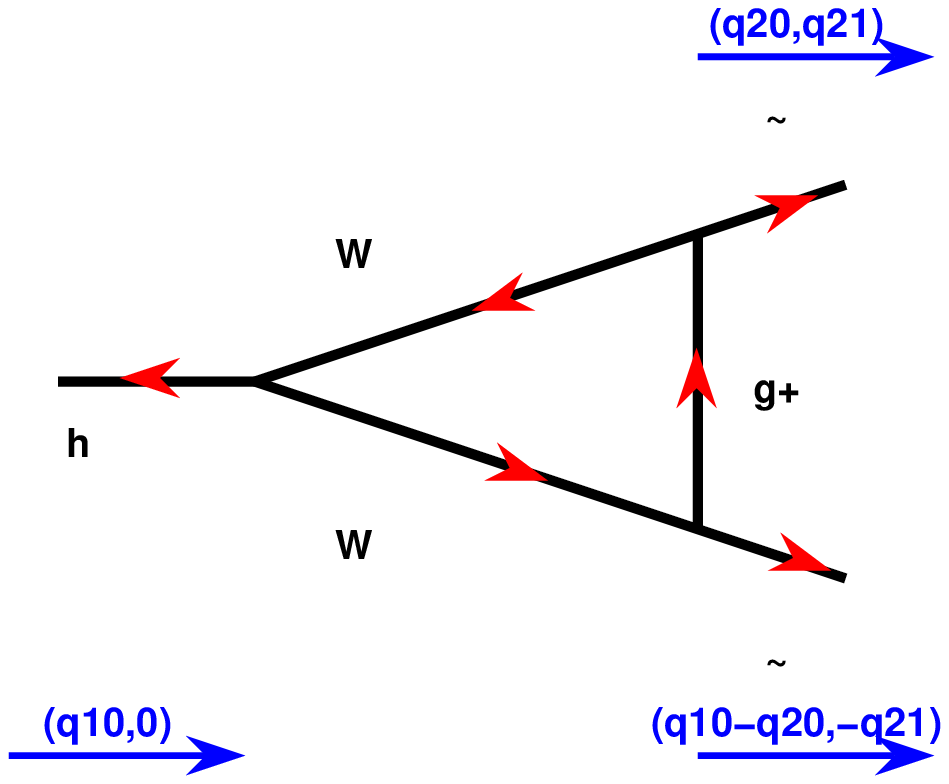,width=3.7cm}
\epsfig{file=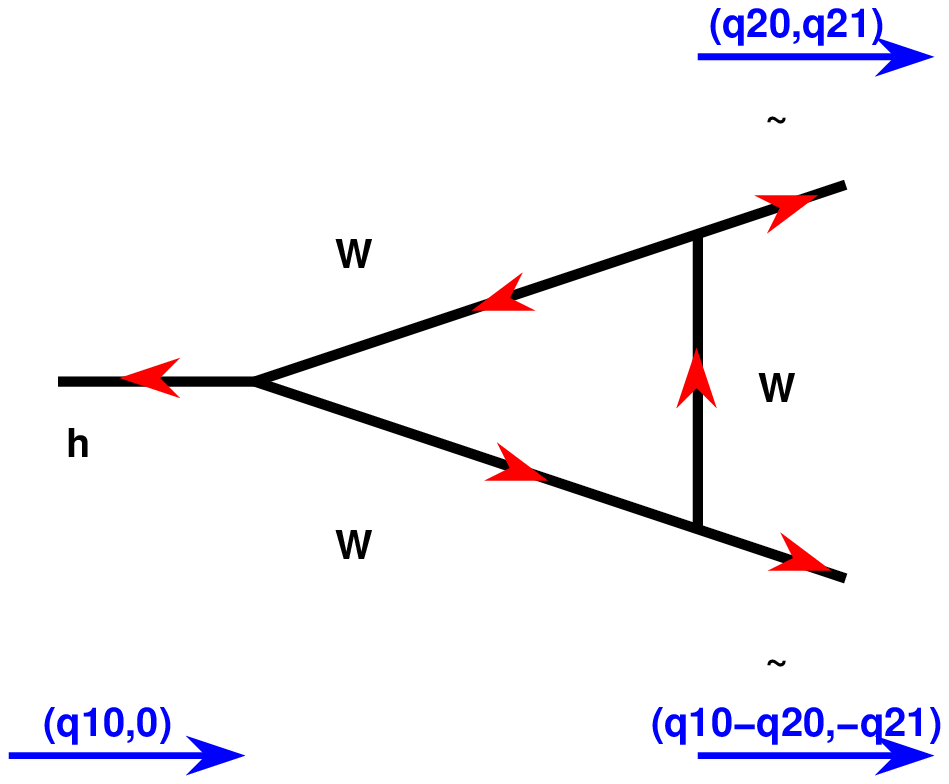,width=3.7cm}
\epsfig{file=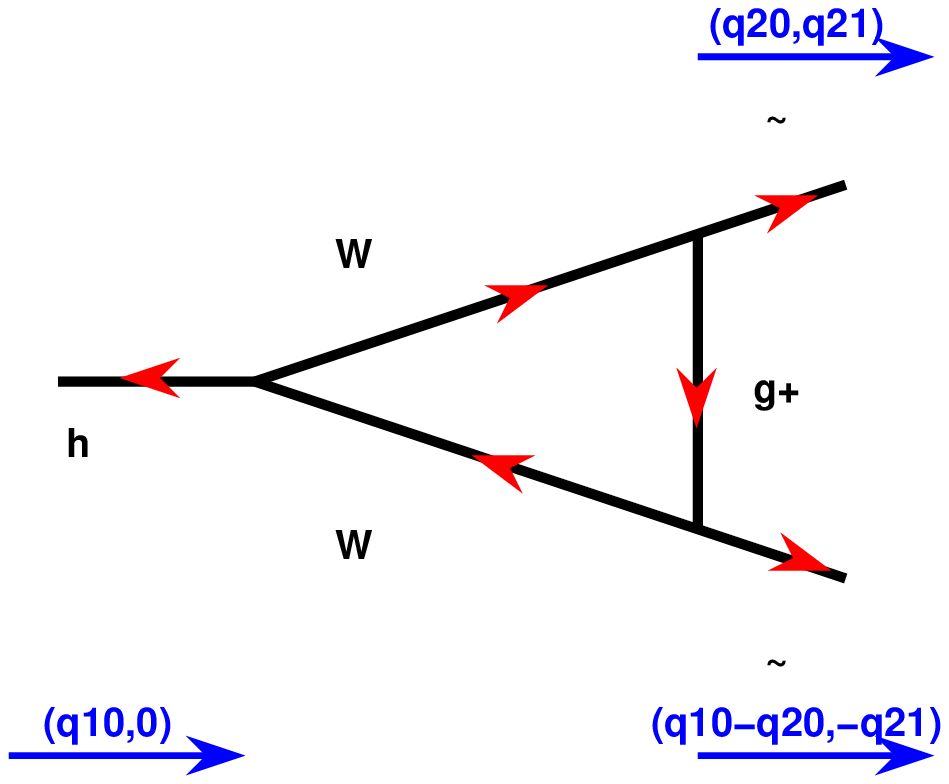,width=3.7cm}\\[0.1cm]
\epsfig{file=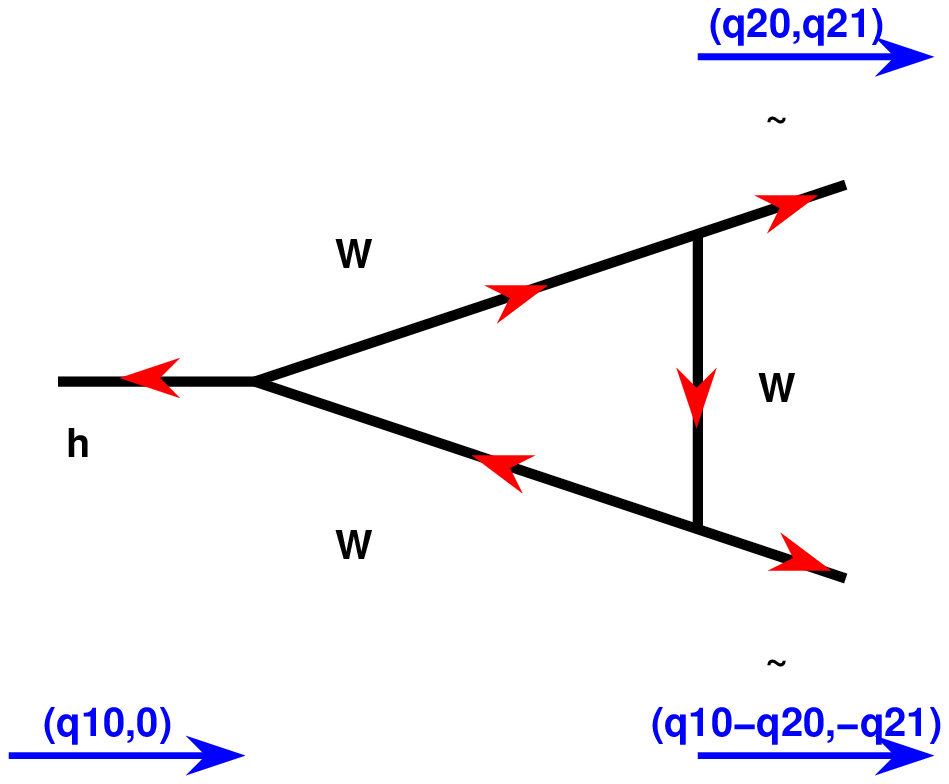,width=3.7cm}
\epsfig{file=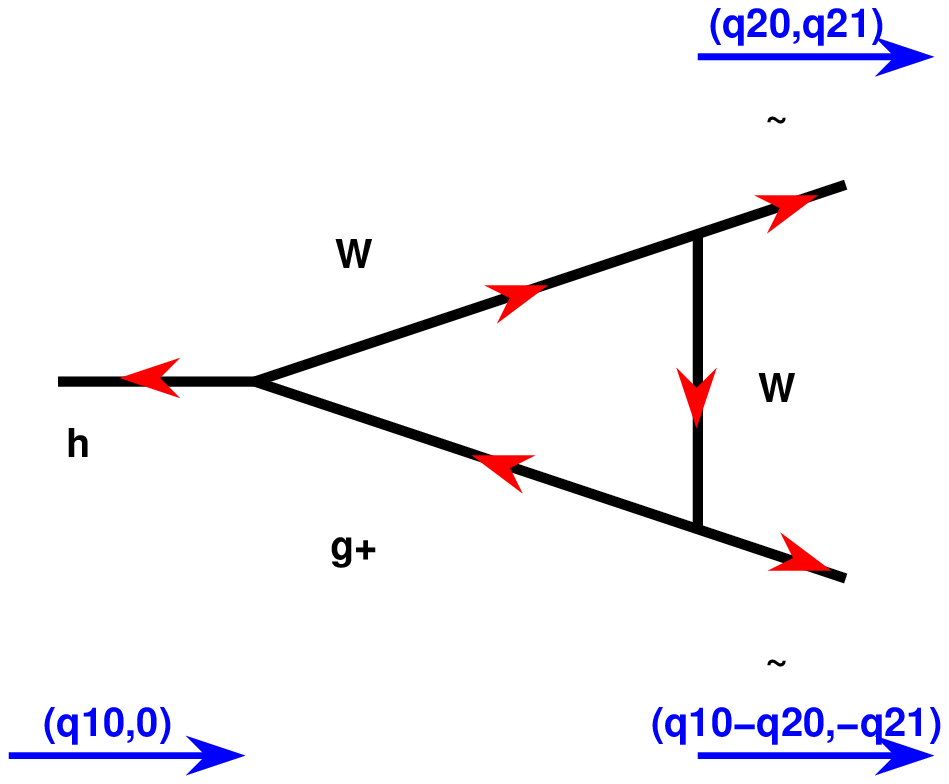,width=3.7cm}
\epsfig{file=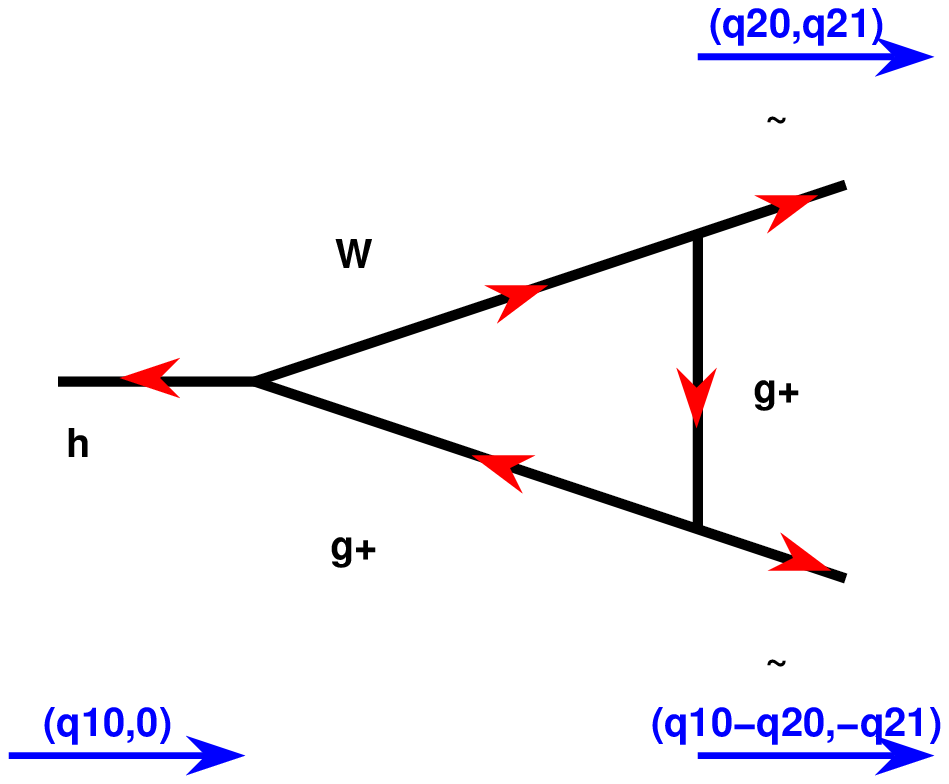,width=3.7cm}
\epsfig{file=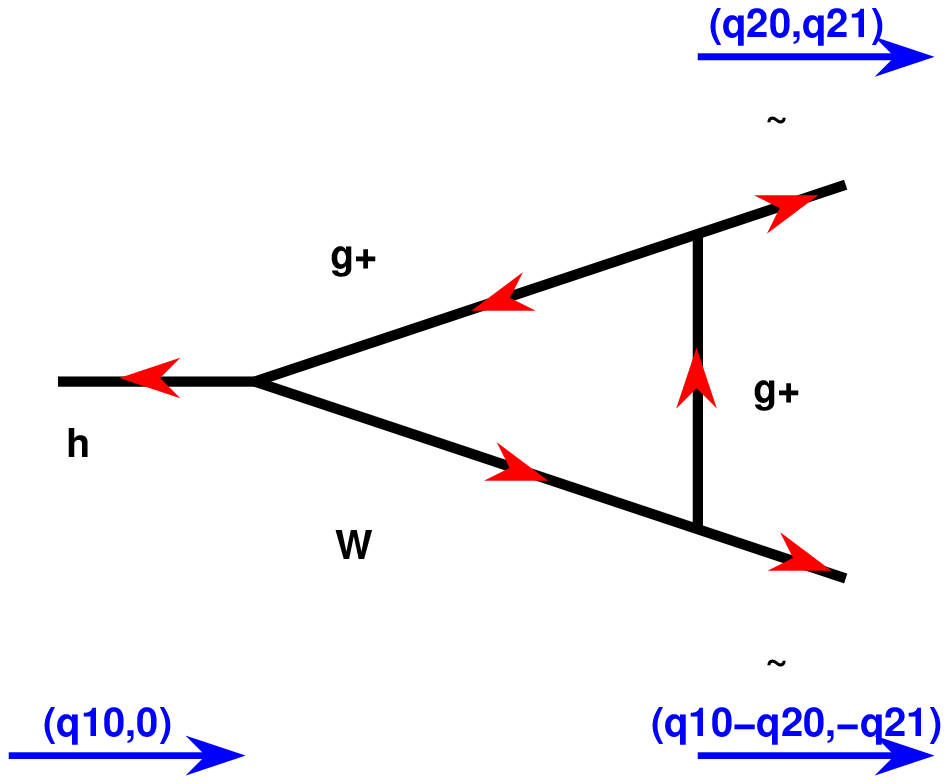,width=3.7cm}\\[0.1cm]
\epsfig{file=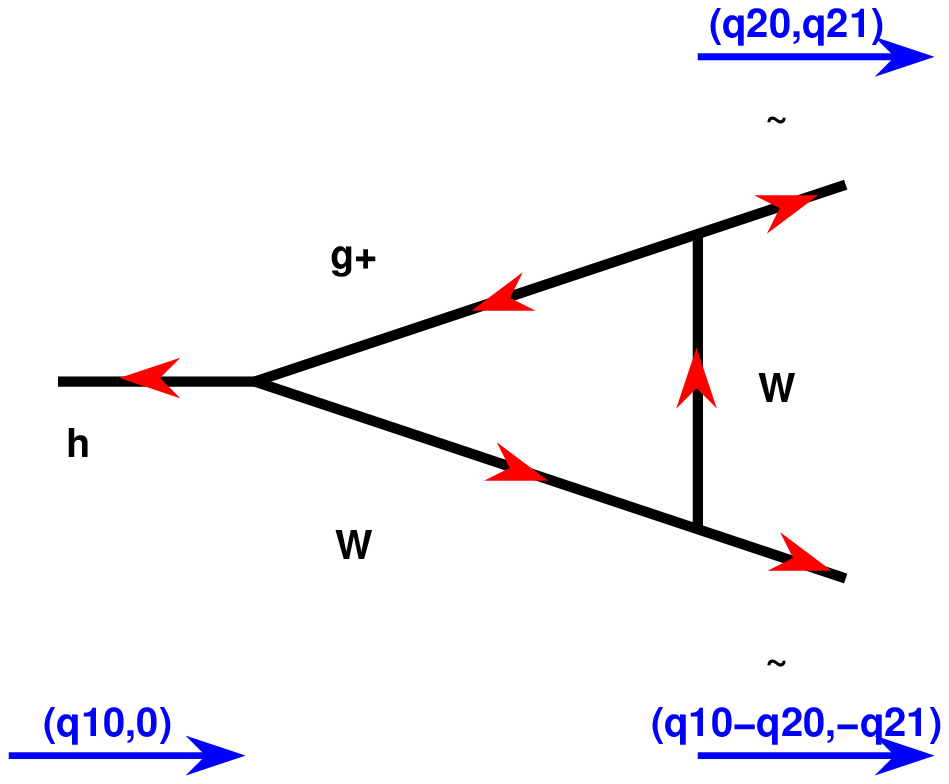,width=3.7cm}
\epsfig{file=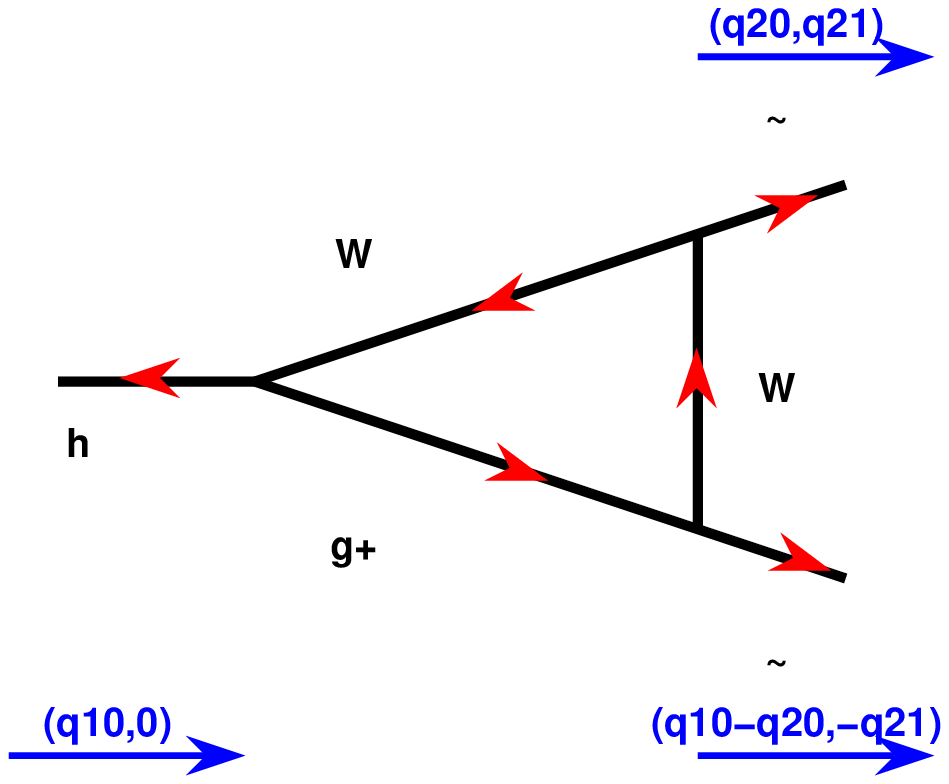,width=3.7cm}
\epsfig{file=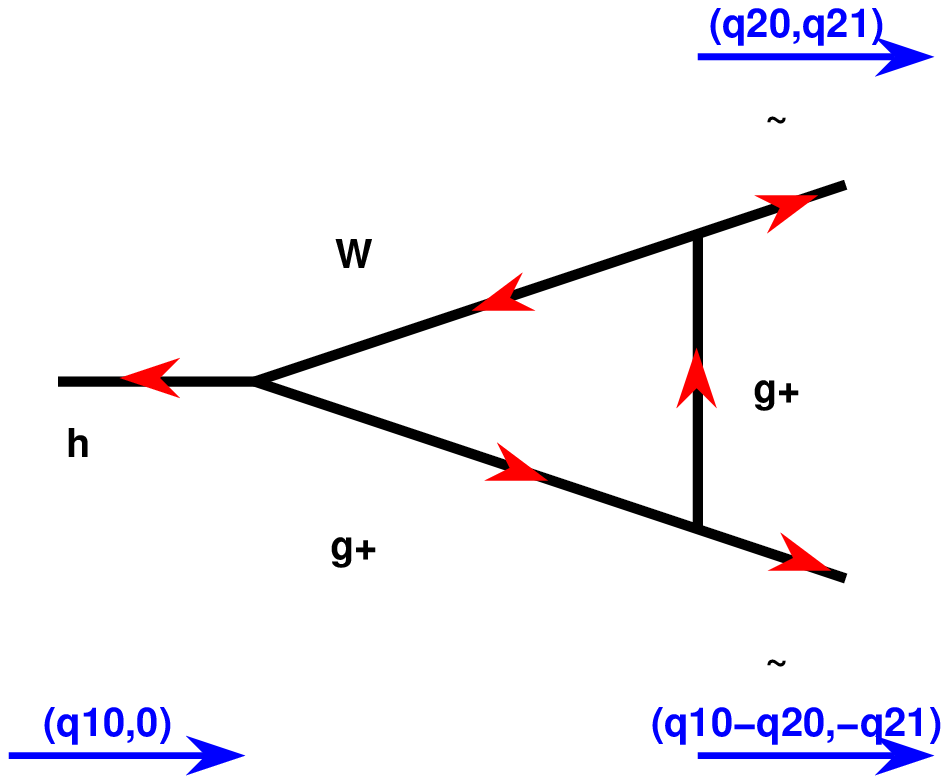,width=3.7cm}
\epsfig{file=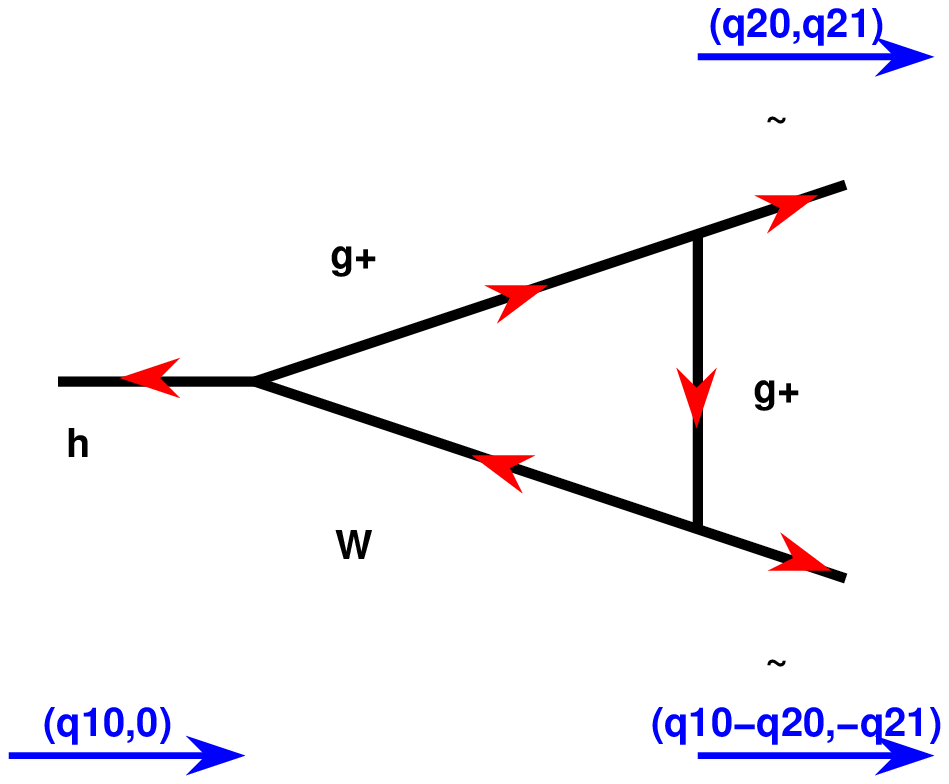,width=3.7cm}\\[0.1cm]
\epsfig{file=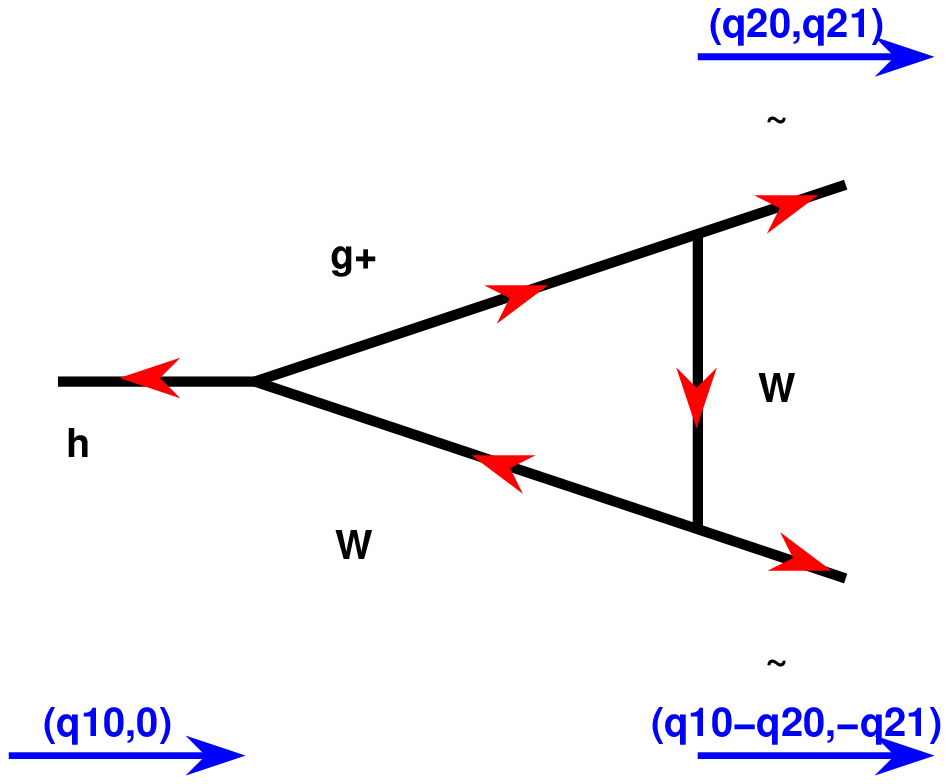,width=3.7cm}
\epsfig{file=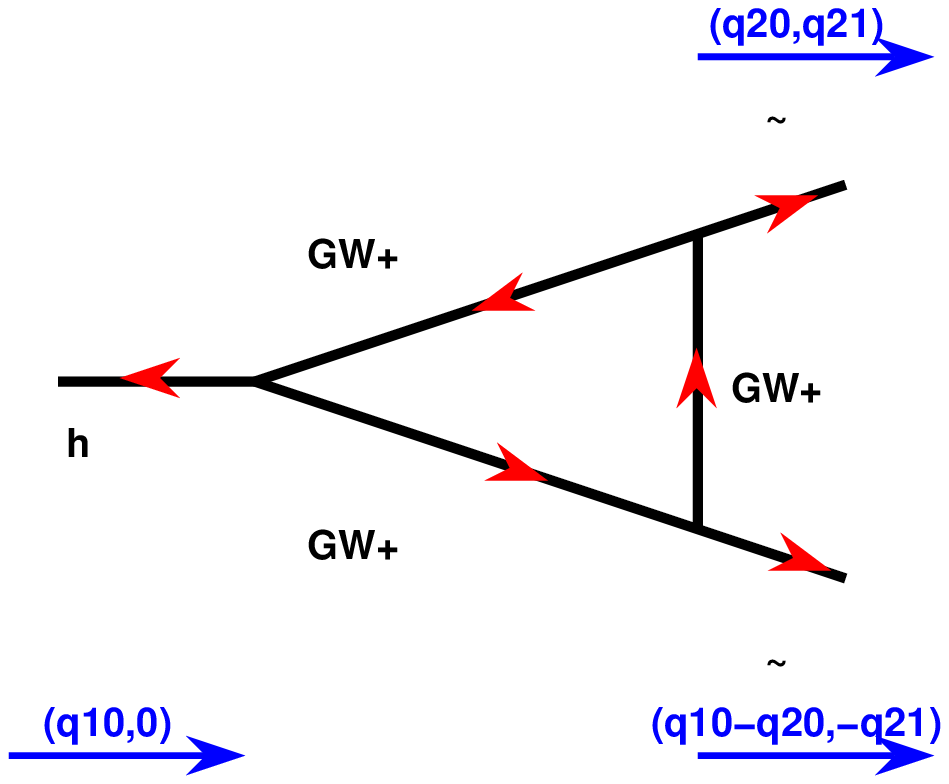,width=3.7cm}
\epsfig{file=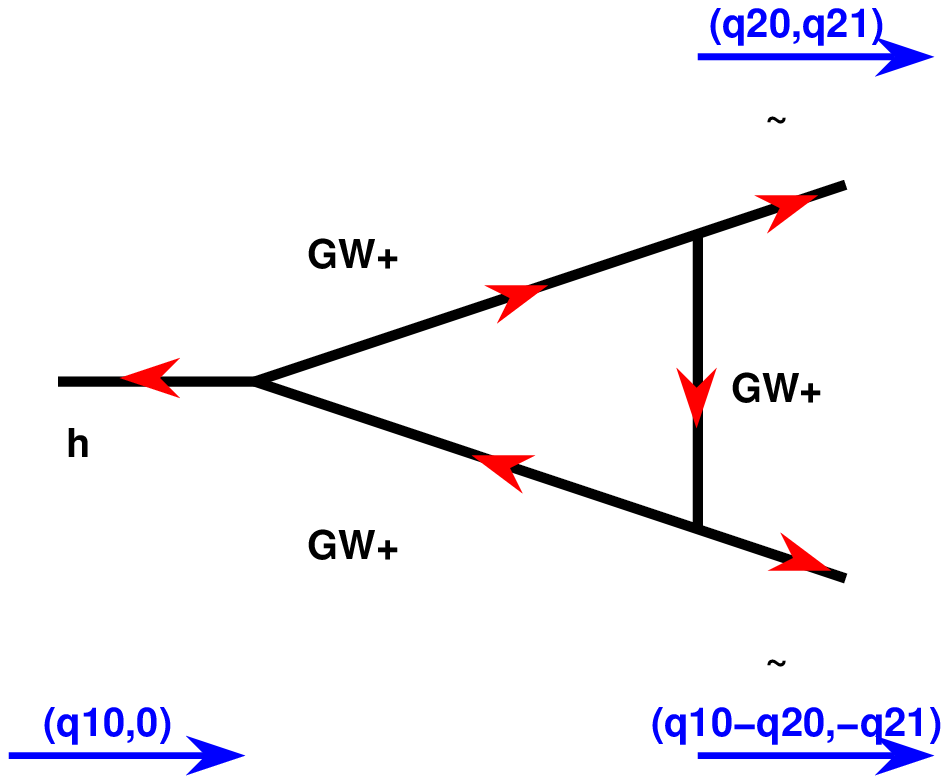,width=3.7cm}
\epsfig{file=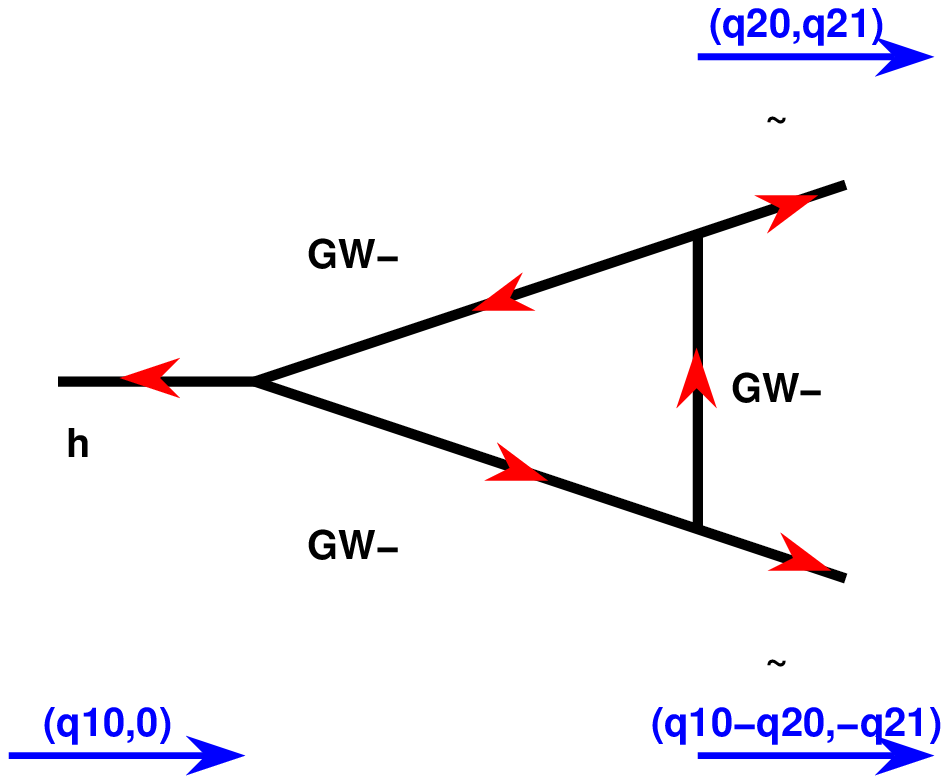,width=3.7cm}\\[0.1cm]
\epsfig{file=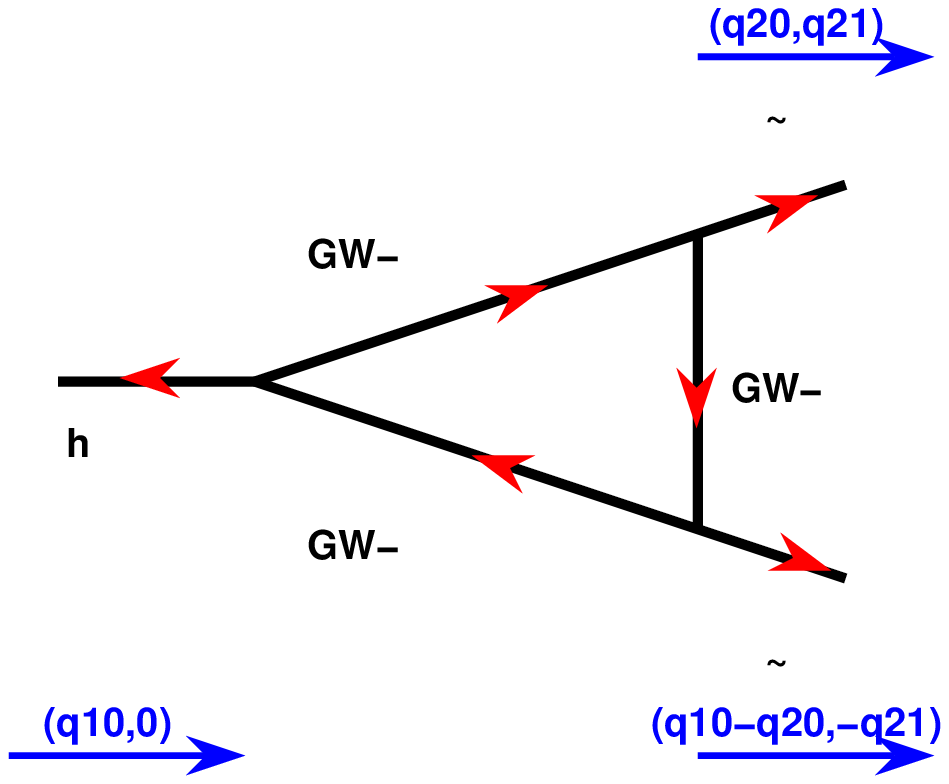,width=3.7cm}\\[0.1cm]
\caption{The contributing graphs to $h^0 \rightarrow \gamma \gamma$ in the
fermiophobic limit.}
\label{fig:graphs}
\end{center}
\end{figure}

In Fig. \ref{fig:graphs} we show all the diagrams that were included. A
previous work by Diaz and Weiler \cite{DW94} did not include the
Higgs-bosons diagrams. Our calculation, in the 'tHooft-Feynman gauge, was
done with {\em xloops}\cite{xl1,xl2}. We have been using this program to
calculate other amplitudes in the framework of the 2HDM \cite{Sant2}.
Throughout this process we have made several checks on the computer results.
In this particular case we have verified that the contribution of the vector
boson loops agrees with a calculation done by M. Spira et al. \cite{spira}
using the supersymmetric version of the 2HDM.

\begin{figure}[htbp]
    \epsfig{file=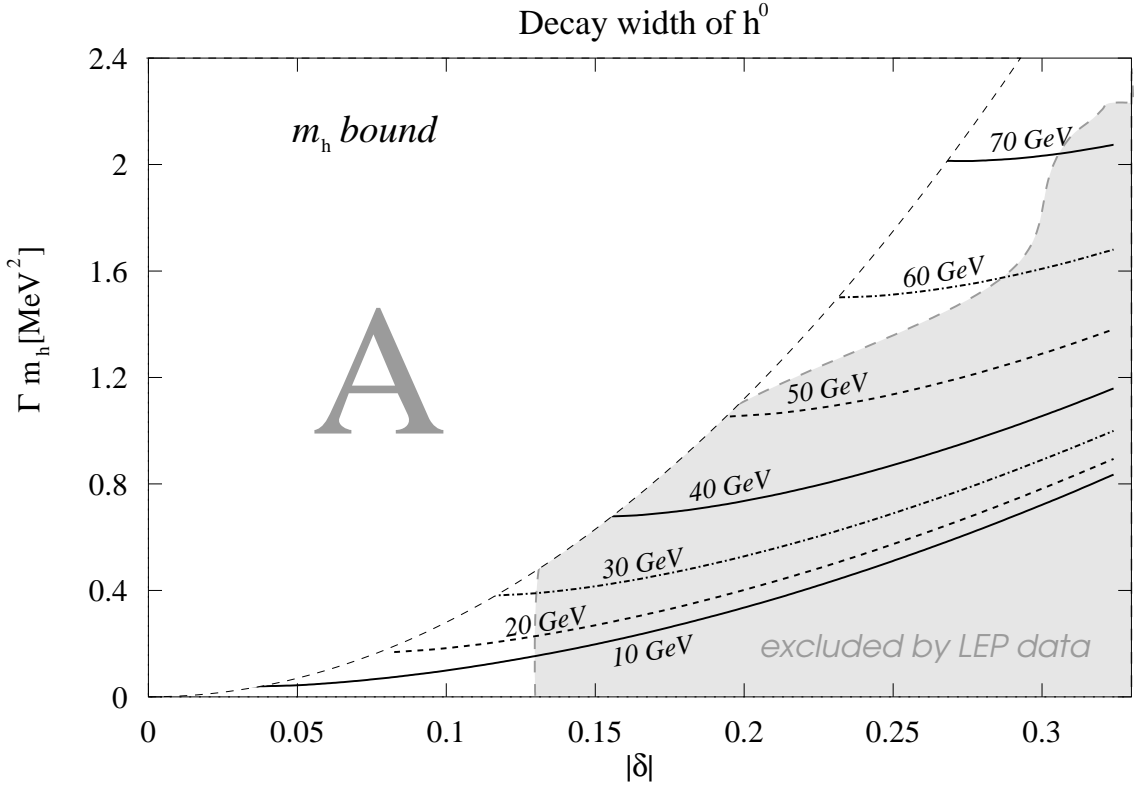,width=15cm}
\caption{Dependence on $\delta$ and $m_{h^0}$ at $m_{H^+}=131 \;GeV$ and $%
\alpha=\frac{\pi}{2}$ in potential A.}
\label{fig:deltmh0_A}
\end{figure}

\begin{figure}[htbp]
    \epsfig{file=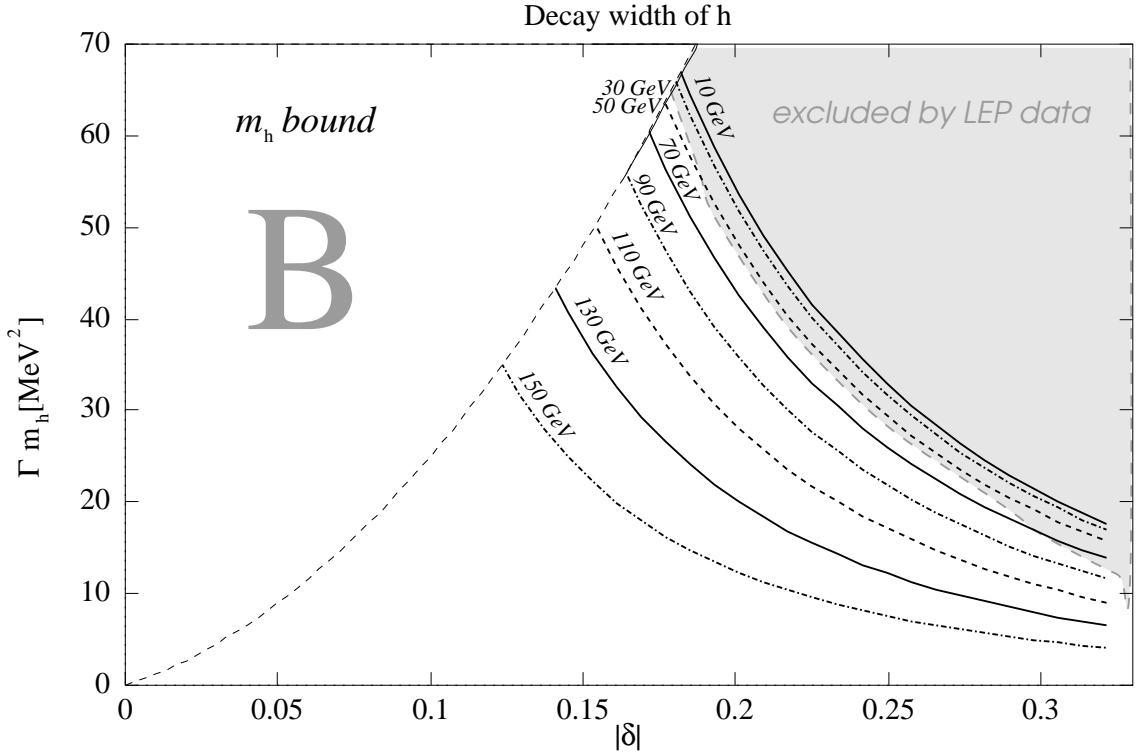,width=15cm}
\caption{Dependence on $\delta$ and $m_{h^0}$ at $m_{H^+}=131 \;GeV$ and $%
\alpha=\frac{\pi}{2}$ in potential B.}
\label{fig:deltmh0_B}
\end{figure}

In Fig. \ref{fig:deltmh0_A} we show the product $m_{h^{0}}$ times the decay
width ($\Gamma $) for the process $h^{0}\rightarrow \gamma \gamma $ in model
A as a function of $\delta $ and for several values of $m_{h^{0}}$ and a
fixed value of $m_{H^{+}}$. This function shows a gentle rise with $m_{h^{0}}
$ which reflects the proportionality between $g_{h^{0}H^{+}H^{-}}$ and $%
m_{h^{0}}^{2}$. Looking at this coupling constant one could naively assume
that there would be an enhancement for $\beta $ approaching $\pi /2$, i. e.,
in our plot, when $\delta $ approaches zero. However, a close examination
shows that such an enhancement does not exist. On the contrary, the coupling
vanishes in this limit, since $m_{h^{0}}$ goes to zero when $\beta
\rightarrow \pi /2$. Alternatively, if one keeps $m_{h^{0}}$ fixed, then the
mass relation 
\begin{equation}
m_{h^{0}}=\sqrt{2\lambda _{1}}v_{1}=\sqrt{2\lambda _{1}}v\cos \beta 
\end{equation}
imposes a lower bound for $\beta $. In Fig.~\ref{fig:deltmh0_A} the dotted line gives this
limit, evaluated assuming  $\lambda _{1}=1/2$. The dashed area shows
the exclusion region implied by the LEP experimental results. In the
work of Ackerstaff et al. \cite{Acker} an experimental bound on the SM
$h \rightarrow \gamma \gamma$ branching ratio is derived. For a
fermiophobic Higgs with $m_h < m_W$ the $\gamma \gamma$ branching
ratio is one. On the other hand, the production mechanism is
suppressed by a factor $\sin^2 \delta$. Hence, we have turned the OPAL
experimental bounds into a bound on $\delta$. Fig.~\ref{fig:deltmh0_B}
gives the equivalent information for potential B.

\begin{figure}[htbp]
    \epsfig{file=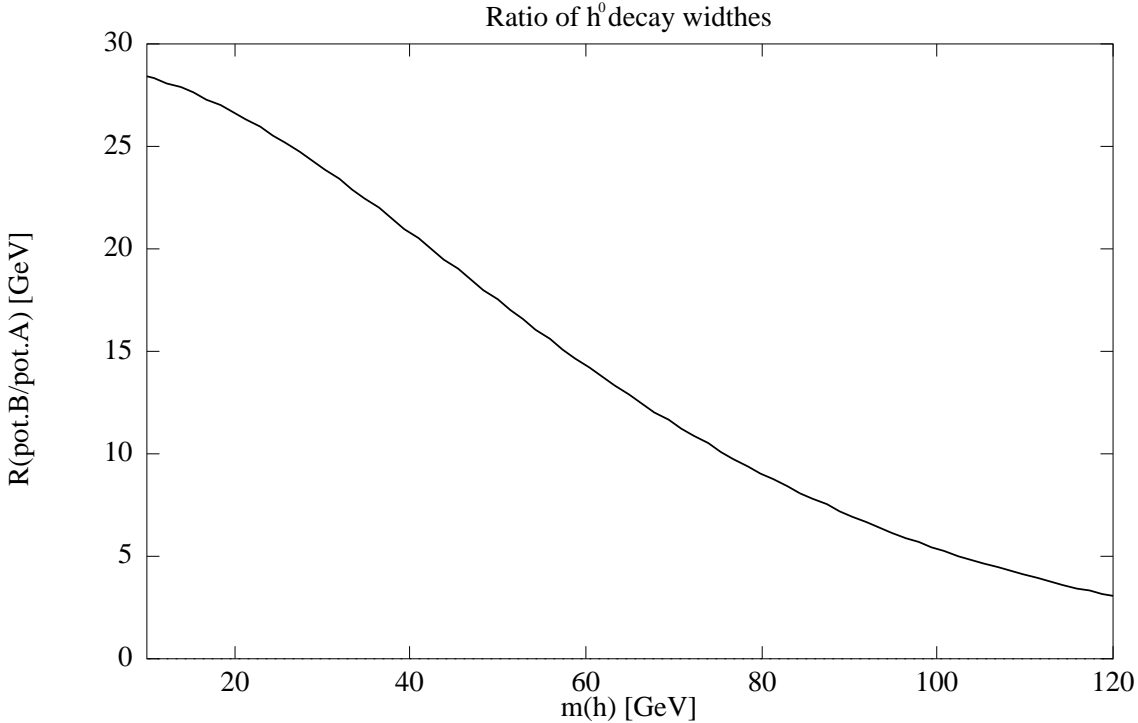,width=15cm}
\caption{Ratio of the decay widthes from $V_{(B)}/V_{(A)}$ with $\delta=0.29$
and $m_{H^+}=200 \;GeV$ and $m_{A^0}=250\;GeV$.}
\label{fig:ratio250}
\end{figure}

\begin{figure}[htbp]
    \epsfig{file=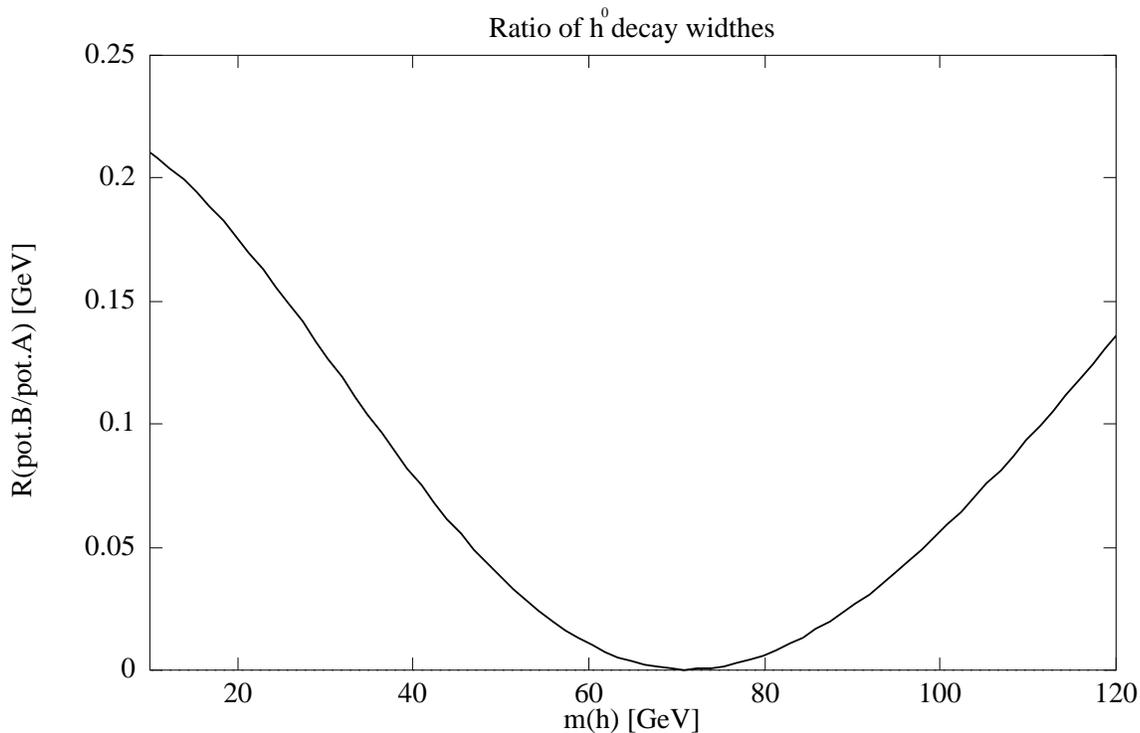,width=15cm}
\caption{Ratio of the decay widthes from $V_{(B)}/V_{(A)}$ with $\delta=0.29$
and $m_{H^+}=200 \;GeV$ and $m_{A^0}=120\;GeV$.}
\label{fig:ratio120}
\end{figure}

In Fig. \ref{fig:ratio250} we plot, as a function of $m_{h^0}$, the ratio $R$%
, of the widthes calculated with potentials $V_B$ and $V_A$, respectively.
According to the fermiophobic limit, we set $\alpha = \pi/2$ and $\delta=0.29
$. For the other relevant masses we have used $m_{H^+}=200 \;GeV/c^2$ and $%
m_{A^0}=250\;GeV/c^2$. In the range of variation of $m_{h^0}$, i.e., $20
\;GeV/c^2 < m_{h^0} < 120 \;GeV/c^2$, R decreases smoothly from 25 till 3.
However, it is misleading to assume that potential A always gives smaller
results. This is clearly shown in Fig. \ref{fig:ratio120} where we plot the
same function $R$ evaluated with the same parameters except for $m_{A^0}$
that was set up to $120\;GeV/c^2$. Again, $R$ is a decreasing function of $%
m_{h^0}$ that has a zero for $m_{h^0}$ around $70\;GeV/c^2$ and increases
afterwards. However, in this case, the values obtained with potential B are
smaller than the corresponding ones for potential A.

This behavior can be qualitatively understood if one examines the coupling
constants $[h^0 H^+ H^-]_{(A)}$ and $[h^0 H^+ H^-]_{(B)}$ given by equations
(\ref{coup1}) and (\ref{coup2}), respectively. In the range of $m_{h^0}$
that we are considering, and for the same values of $\alpha$, $\beta$ and $%
m_{H^+}$ , the coupling corresponding to potential $A$ is always negative
and decreases from about $- 85 \;GeV/c^2$ till $- 230 \;GeV/c^2$. On the
contrary, the coupling constant corresponding to potential $B$ is positive.
For large values of $m_{A^0}$ (around $250 \;GeV/c^2$), it decreases from $%
930 \;GeV/c^2$ till $780 \;GeV/c^2$ for $20 \;GeV/c^2 < m_{h^0} < 120
\;GeV/c^2$. These values of the coupling constant, when compared with the
corresponding ones for potential $A$, explain the qualitative behavior of
the ratio $R$ given in Fig. \ref{fig:ratio250}. The explanation of
Fig. \ref{fig:ratio120} is more subtle, but 
again, it depends on the coupling constant of potential $B$. In fact, when $%
m_{A^0}=120\;GeV/c^2$ the coupling corresponding to potential $B$ starts at $%
100\;GeV/c^2$ and decreases smoothly till $-60\;GeV/c^2$, having a zero
around $m_{h^0}=95\;GeV/c^2$. This behavior has two consequences. When the
coupling is positive, its order of magnitude is the correct one to almost
cancel the W-loops contributions to the width. Hence, $R$ is small because
potential $B$ gives a small width. This cancellation is exact for $m_{h^0}$
around $70\;GeV/c^2$ and after that, because the coupling changes sign, the
charged Higgs contribution adds up to the normal W-loop result. Hence $R$
increases.

Despite the fact that the $[hWW]$ coupling is suppresed by $\sin
\delta$, one should keep in mind that when $m_h$ is larger then $m_W$
the decay channel $h\rightarrow W W^* \rightarrow W q \bar{q}$ starts
to compete with the $\gamma\gamma$ channel. We have evaluated the
$WW^*$ decay width and in table \ref{tab1} we show some results in
comparison with the width for the $\gamma \gamma$ channel evaluated for
potential A and $m_{H^+}=100\ GeV$. The table is representative of a
situation that can be summarized qualitatively as follows:
i) for small $\delta \ (\delta = 0.1)$ the $WW^*$ width is comparable
with the $\gamma\gamma$ width for $m_h=120\ GeV/c^2$;  
ii) for large $\delta \ (\delta = 0.3)$ even at $m_h = 120 \ GeV/c^2$
the $WW^*$ decay width is already larger than the $\gamma\gamma$ width
by a factor of ten.

\begin{table}[h]
  \begin{center}
    \begin{tabular}[b]{|r|cc|cc|}
  & \multicolumn{2}{c|}{$\delta = 0.1$} & \multicolumn{2}{c|}{$\delta = 0.3$}  \\
  \phantom{2345}$m_h$ & $h\rightarrow WW^*$ & $h\rightarrow \gamma\gamma$(A) &
  $h\rightarrow WW^*$ & $h\rightarrow \gamma\gamma$(A) \\
\hline
  $90$ & $2.6\times 10^{-6}$ & $0.4\times 10^{-3}$ & $2.3\times 10^{-5}$ & $0.6\times 10^{-3}$ \\ 
  $120$ & $2.2\times 10^{-3}$ & $2.7\times 10^{-3}$ & $1.9\times 10^{-2}$ & $2.2\times 10^{-3}$ \\ 
  $150$ & $5.7\times 10^{-2}$ & $1.5\times 10^{-2}$ & $5.0\times 10^{-1}$ & $8.0\times 10^{-3}$  
   \end{tabular}
    \caption{Comparism between the widthes for the $WW^*$ and $\gamma\gamma$ channels.}
    \label{tab1}
  \end{center}
\end{table}

\section{Conclusion}

We have examined the 2HDM where the potential does not explicitly break CP
violation  and furthermore it is naturally protected from the appearance of
minima with CP violation \cite{Sant3}. There are two ways of accomplishing
this, leading to two different potentials $V_A$ and $V_B$. $V_A$ is
invariant under the discrete group $Z_2$ and $V_B$ is invariant under $U(1)$
except for the presence of a soft breaking term. These two symmetries ensure
that the parameters that, at tree-level, were set to zero, are not required
to renormalize the models.

The potential $V_A$ and $V_B$ have different cubic and quartic scalar
vertices. Then, it is obvious that they give different Higgs-Higgs
interactions. However, even before one is able to test such interactions,
one could still sense these two different phenomenologies via Higgs-loop
contributions.

To illustrate this point we have considered a fermiophobic neutral Higgs,
decaying mainly into two photons. The widthes for the decays calculated with
both potentials can differ by orders of magnitude for reasonable values of
the parameters. Clearly, with four masses and two angles as free parameters,
it is not worthwhile to perform a complete analysis. Nevertheless, we
believe that the results presented here are sufficient for illustrative
purposes. The experimental searches in this area should be made with an open
mind for surprises.

\section{Acknowledgment}

We would like to thank our experimental colleagues at LIP for some useful
discussions and the theoretical elementary particle physics department of
Mainz University for allowing us to use their computer cluster. L.B. is
partially supported by JNICT contract No. BPD.16372.


\begin{references}
\bibitem{vanc}  {M. Martinez {\it et al.}}.{\em Precision Tests of the
Electroweak Interactions at the Z pole}, CERN-EP/98-27.

\bibitem{Sant3}  {J. Velhinho, R. Santos, A. Barroso}. {\em Phys. Lett.} 
{\bf B 322} (1994) 213--218.

\bibitem{sher1}  {M.Sher}.{\em Phys. Rep.}{\bf 179} (1989) 273.

\bibitem{Gun}  {J. F. Gunion, H. E. Haber, G. Kane, S. Dawson}. {\ {\em The
Higgs Hunter's Guide}}. Addison Wesley (1990)

\bibitem{Sant1}  {R. Santos, A. Barroso}. {\em Phys. Rev.} {\bf D 56} (1997)
5366.

\bibitem{lepweg}  {LEPWEG, The Lep Collaborations ALEPH, DELPHI, L3, OPAL,
The LEP Electroweak Working Group and the SLD Heavy Flavour Group, 1997.}%
{\em A Combination of Preliminary Electroweak Measurements and Constraints
on the Standard Model}, CERN--PPE/97-154.

\bibitem{kraw1}  {M. Krawczyk, J. Zochowski, P. M\"{a}ttig}. {\em Eprint} {\
hep-ph/9811256} {\em and references therein.}

\bibitem{Aker1}  {A. G. Akeroyd}, {\em Eprint} {\ hep-ph/9806337.} \newline
{S. Nie and M. Sher}, {\em Eprint} {\ hep-ph/9811234.}

\bibitem{ciu1}  {M. Ciuchini, G. Degrassi, P. Gambino, G.F. Giudice}.{\em %
Nucl. Phys. } {\bf B 527} (1998) 21.

\bibitem{cleo1}  {M. S. Alam {\it et al.} (CLEO Colla.)}, {\em Phys. Rev.
Lett.} {\bf 74} (1995) 2885.

\bibitem{den1}  {A. Denner, R.J. Guth, W. Hollik, J.H. K\"{u}hn}.{\em Z.
Phys.} {\bf C 51} (1991) 695-705. \newline
{S. Bertolini}, {\em Nucl. Phys.} {\bf B272} (1986) 77.

\bibitem{Rev}  {Review of particle properties}, {\em Phys. Rev.} {\bf D54}
(1996), 1.

\bibitem{Lang98}  Talk given at the 5th International Wein Symposium
(WEIN98). Santa Fe, 1998. Eprint: hep-ph/ 9809352.

\bibitem{DW94}  M. Diaz and T. Weiler, unpublished, hep-ph/9401259.

\bibitem{xl1}  {L. Br\"{u}cher, J. Franzkowski, D. Kreimer}. {\em Nucl.
Instrum. Meth. } {\bf A 389} (1997) 323--342.

\bibitem{xl2}  {L. Br\"{u}cher, J. Franzkowski, D. Kreimer}. {\em Eprint} {\
hep-ph/9710484}.

\bibitem{Sant2}  {A. Barroso, L. Br\"{u}cher, R. Santos}. {\em Phys.Lett.} 
{\bf B 391} (1997) 429--433.

\bibitem{spira}  {M. Spira, A. Djouadi, D. Graudenz, P. M. Zerwas}. {\em %
Nucl. Phys.} {\bf B453} (1995) 17.


\bibitem{Acker}
K.~Ackerstaff {\it et al.}
[OPAL Collaboration]. Phys. Lett. {\bf B437} (1998) 218. 

\end{references}

\end{document}